\documentclass[12pt,draftcls,onecolumn]{IEEEtran}
\usepackage{setspace}
\usepackage{amsmath, amsfonts, amssymb}
\usepackage{algorithmic}
\usepackage{algorithm}
\usepackage{array}
\usepackage[caption=false,font=normalsize,labelfont=scriptsize,textfont=scriptsize]{subfig}
\usepackage{textcomp}
\usepackage{stfloats}
\usepackage{url}
\usepackage{verbatim}
\usepackage{graphicx}
\usepackage{cite}
\usepackage{multirow}
\usepackage{dashrule}
\usepackage{xcolor}
\newtheorem{remark}{Remark}
\newtheorem{theorem}{Theorem}
\newtheorem{lemma}{Lemma}
\begin{document}

\title{\huge Two-Dimensional Direction-of-Arrival Estimation Using Stacked Intelligent Metasurfaces}
\author{Jiancheng An,~\IEEEmembership{Member,~IEEE}, Chau Yuen,~\IEEEmembership{Fellow,~IEEE},\\Yong Liang Guan,~\IEEEmembership{Senior Member,~IEEE}, Marco Di Renzo,~\IEEEmembership{Fellow,~IEEE},\\M\'erouane Debbah,~\IEEEmembership{Fellow,~IEEE}, H. Vincent Poor,~\IEEEmembership{Life Fellow,~IEEE},\\and Lajos Hanzo,~\IEEEmembership{Life Fellow,~IEEE}
\thanks{J. An, C. Yuen, and Y. L. Guan are with the School of Electrical and Electronics Engineering, Nanyang Technological University, Singapore 639798 (e-mail: jiancheng.an@ntu.edu.sg; chau.yuen@ntu.edu.sg; eylguan@ntu.edu.sg).}
\thanks{M. Di Renzo is with CNRS, CentraleSup\'elec, Laboratoire des Signaux et Syst\`emes, Universit\'e Paris-Saclay, 91192 Gif-sur-Yvette, France (e-mail: marco.di-renzo@universite-paris-saclay.fr).}
\thanks{M. Debbah is with the Center for 6G Technology, Khalifa University of Science and Technology, P O Box 127788, Abu Dhabi, United Arab Emirates (e-mail: merouane.debbah@ku.ac.ae).}
\thanks{H. Vincent Poor is with the Department of Electrical and Computer Engineering, Princeton University, Princeton, NJ 08544 USA (e-mail: poor@princeton.edu).}
\thanks{L. Hanzo is with the School of Electronics and Computer Science, University of Southampton, SO17 1BJ Southampton, U.K. (e-mail: lh@ecs.soton.ac.uk).}\vspace{-0.8cm}}
\markboth{DRAFT}{DRAFT}

\maketitle

\begin{abstract}
Stacked intelligent metasurfaces (SIM) are capable of emulating reconfigurable physical neural networks by relying on electromagnetic (EM) waves as carriers. They can also perform various complex computational and signal processing tasks. A SIM is fabricated by densely integrating multiple metasurface layers, each consisting of a large number of small meta-atoms that can control the EM waves passing through it. In this paper, we harness a SIM for two-dimensional (2D) direction-of-arrival (DOA) estimation. In contrast to the conventional designs, an advanced SIM in front of the receiver array automatically carries out the 2D discrete Fourier transform (DFT) as the incident waves propagate through it. As a result, the receiver array directly observes the angular spectrum of the incoming signal. In this context, the DOA estimates can be readily obtained by using probes to detect the energy distribution on the receiver array. This avoids the need for power-thirsty radio frequency (RF) chains. To enable SIM to perform the 2D DFT, we formulate the optimization problem of minimizing the fitting error between the SIM's EM response and the 2D DFT matrix. Furthermore, a gradient descent algorithm is customized for iteratively updating the phase shift of each meta-atom in SIM. To further improve the DOA estimation accuracy, we configure the phase shift pattern in the zeroth layer of the SIM to generate a set of 2D DFT matrices associated with orthogonal spatial frequency bins. Additionally, we analytically evaluate the performance of the proposed SIM-based DOA estimator by deriving a tight upper bound for the mean square error (MSE). Our numerical simulations verify the capability of a well-trained SIM to perform DOA estimation and corroborate our theoretical analysis. It is demonstrated that a SIM having an optical computational speed achieves an MSE of $10^{-4}$ for DOA estimation.
\end{abstract}


\section{Introduction} 
\IEEEPARstart{D}{irection-of-arrival} (DOA) estimation using multiple antenna technology has long been a crucial task with compelling applications in areas such as astronomy, navigation, and the emerging integrated sensing and communication systems of the sixth-generation (6G) networks \cite{arXiv_2023_An_Stacked_DOA, TAP_1986_Schmidt_Multiple, TASSP_1989_Roy_ESPRIT, TASSP_1989_Stoica_MUSIC, TASSP_1989_Rao_Performance, TSP_2015_Qin_Generalized, TSP_2013_Yang_Off, TWC_2023_An_Fundamental, SPL_2020_Zheng_Direction}. Traditionally, DOAs have been estimated using the classic beamforming method, which can be efficiently implemented via the fast Fourier transform technique \cite{BOOK_1985_Haykin_Array, SPM_1996_Krim_Two}. However, the angular resolution of this method is fundamentally restricted by the array aperture \cite{Proc_1997_Godara_Application}. The Rayleigh criterion indicates that a pair of signal sources can only be distinguished, when their angular separation exceeds the antenna beamwidth. To address this problem, several super-resolution DOA estimation approaches have been developed \cite{TAP_1986_Schmidt_Multiple, TASSP_1989_Roy_ESPRIT, TASSP_1989_Stoica_MUSIC, TASSP_1989_Rao_Performance}. The two most prominent techniques are the multiple signal classification (MUSIC) \cite{TAP_1986_Schmidt_Multiple} and the estimation of signal parameters via rotational invariance techniques (ESPRIT) \cite{TASSP_1989_Roy_ESPRIT}. Specifically, MUSIC leverages the orthogonality of the signal and noise subspaces to construct a spatial spectrum \cite{TAP_1986_Schmidt_Multiple}, and then the DOA parameters of the received signals are identified via spectral peak search. By contrast, ESPRIT exploits the spatial rotational invariance property of the signal subspace \cite{TASSP_1989_Roy_ESPRIT}, avoiding spectral search and substantially improving the computational efficiency. Since then, a variety of modifications of MUSIC and ESPRIT have been proposed for different scenarios \cite{TASSP_1989_Rao_Performance}.

While these methods provide significant performance advantages, they require increased computational and storage resources for performing eigenvalue decomposition of the spatial covariance matrix. However, their performance degrades noticeably, when only a small number of snapshots are available. Moreover, traditional DOA estimation methods assume ideal signal conditions and perfect antenna arrays, which cannot be satisfied in practical systems. Various imperfections like non-ideal transceiver design, station location errors, and background radiation significantly degrade the performance of parametric methods, which are impervious to modeling and calibration using conventional techniques \cite{arXiv_2023_An_Toward, TSP_2004_See_Direction, TSP_2013_Liu_A}.

Fortunately, advanced machine learning (ML) techniques provide effective approaches for estimating the DOAs. In contrast to model-based conventional methods, ML-based approaches are data-driven and thus have the potential to adapt to complex electromagnetic (EM) environments and be more robust against practical array imperfections. For instance, the authors of \cite{TAP_2018_Liu_Direction} introduced a deep neural network (DNN) based framework having two parts. A multitasking autoencoder first decomposed the input signals into multiple components falling into distinct angular intervals. Then the results of multiple parallel classifiers were combined to reconstruct the angular-domain spectrum and to estimate the signal directions with enhanced robustness. Furthermore, the authors of \cite{JSTSP_2019_Chakrabarty_Multi} formulated the DOA estimation of multiple sources as a multi-label classification problem. The phase components of the received signals' short-term Fourier transform coefficients were directly fed into a well-trained convolutional neural network (CNN) to localize speakers in dynamic acoustic scenarios. Additionally, in \cite{SPL_2019_Wu_Deep} a deep convolution network (DCN) was designed for learning the transformation from the undersampled array covariance matrix to the angular spectrum. In contrast to conventional sparse recovery methods requiring complex iterations, the DCN framework has superior computational efficiency. Leveraging the sparsity also improves the DOA estimation accuracy. Nevertheless, DL-based methods require a complex model training phase, resulting in high hardware and computational complexity \cite{TAP_2018_Liu_Direction}. It is also challenging to obtain a large training dataset covering all possible signal distributions, especially in the face of high-Doppler scenarios.

Traditional DNNs rely on commercial processors or dedicated chips to perform computations, whose speed is limited by digital hardware. Recently, a novel diffractive deep neural network (D$^2$NN) was developed using three-dimensional (3D) printed diffractive layers \cite{arXiv_2023_An_Stacked_mag, Science_2018_Lin_All}. D$^2$NN allows large-scale parallel calculations and analog signal processing to be carried out at the speed of light \cite{Science_2014_Silva_Performing}. However, once fabricated, the wave-based D$^2$NN architecture presented in \cite{Science_2018_Lin_All} is fixed, hence limiting its functionality in practice. Motivated by advances in metasurface technologies \cite{JSAC_2020_Di_Smart, TCOM_2022_An_Low, WC_2022_An_Codebook, Light_2014_Cui_Coding, TGCN_2022_An_Joint, TVT_2023_Xu_Channel, IoTJ_2024_Xu_A, TWC_2023_Xu_Antenna}, the authors of \cite{NE_2022_Liu_A} customized a reconfigurable D$^2$NN using stacked intelligent metasurfaces (SIMs). Specifically, a SIM employs an array of programmable metasurface layers, each containing many programmable meta-atoms that can manipulate the EM wave behavior as waves pass through it \cite{JSAC_2023_An_Stacked}. Adapting the bias voltage via a customized field-programmable gate array (FPGA) module allows each meta-atom to act as a reprogrammable artificial neuron having tunable weights.

When EM waves pass through a meta-atom within the SIM, the transmitted wave is determined by multiplying the incoming wave with the meta-atom's complex transmission coefficient \cite{ICC_2023_Nadeem_Hybrid}. According to the Huygens-Fresnel principle \cite{Science_2018_Lin_All}, the wave propagating through each meta-atom acts as a secondary source that illuminates all the meta-atoms on the next layer. All transmitted waves impinging at a neuron on the next layer are superimposed, acting as the corresponding aggregate incident wave. This process continues through each metasurface layer in the SIM. As a consequence, the forward propagation model in the SIM implicitly defines a fully connected artificial neural network (ANN), whose architecture can be reconfigured for realizing sophisticated processing functions \cite{JSAC_2023_An_Stacked, NE_2022_Liu_A}.

Several pioneering efforts have been made using SIMs for performing various signal processing tasks in the EM wave domain. Specifically, the authors of \cite{NE_2022_Liu_A} experimentally evaluated a SIM's capabilities for image classification. They built a SIM prototype having five programmable metasurface layers and used it for recognizing handwritten digits. The first metasurface layer acted as a digital-to-analog converter (DAC), converting each input image to greyscale and configuring its transmission coefficients to match the pixel values. The remaining four layers formed the image recognition neural network. Test on the MNIST dataset demonstrated that the well-trained SIM achieved an accuracy of $90.76\%$ at recognizing the digits $0 \sim 9$. In \cite{JSAC_2023_An_Stacked}, the authors harnessed a SIM for implementing multiple-input multiple-output (MIMO) communications \cite{CL_2023_An_A}. In contrast to conventional MIMO designs, a pair of SIMs deployed at the transmitter and receiver can automatically accomplish MIMO-oriented transmit precoding and receiver combining as the EM waves propagate through them. This allows each spatial stream to be directly radiated and recovered from its corresponding transmit and receive ports, while significantly reducing the number of radio frequency (RF) chains needed. Furthermore, in \cite{ICC_2023_An_Stacked, arXiv_2023_An_Stacked} the authors integrated a SIM into the BS to facilitate downlink multiuser beamforming, while eliminating the conventional digital beamformer and high-resolution DACs at the BS.

\begin{figure*}[!t]
\centering
\includegraphics[width=16cm]{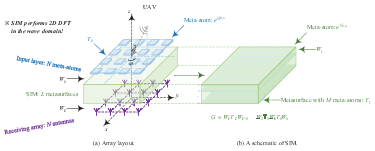}
\caption{A SIM-aided array system.}\vspace{-0.6cm}
\label{fig_2}
\end{figure*}
Nevertheless, DOA estimation using an advanced SIM has hitherto remained unexplored. Hence this is the first paper on this intriguing subject, in which we design a new SIM-based physical DOA estimator. The underlying philosophy is that by appropriately optimizing the SIM to carry out the two-dimensional (2D) discrete Fourier transform (DFT), the incident EM waves can be transformed into the angular frequency domain, as they propagate through the SIM \cite{arXiv_2023_An_Stacked_DOA}. By detecting the energy levels of the different receiver probes -- each corresponding to a unique DOA -- we can read the signal's direction from the probe having the strongest energy. As a result, the receiver hardware is substantially simplified as the analog-to-digital converters (ADC) are no longer needed. Most remarkably, in contrast to previous DOA estimators relying on array signal processing, the calculation within the SIM occurs naturally without incurring extra delay. More specifically, the detailed contributions of this paper are summarized as follows:
\begin{enumerate}
\item We conceive a SIM-based physical estimator to probe the DOA of radio waves impinging from a single source. The SIM is made of multiple metasurface layers, each containing a large number of small meta-atoms having adjustable EM properties. By manipulating the wave propagation therein, the SIM becomes capable of performing the desired signal processing naturally, as the waves pass through each layer, which is significantly faster than traditional digital calculations. A uniform planar array (UPA) is placed at the end of the SIM. By appropriately configuring the SIM, it can transform the radio waves into the angular domain. Each antenna then corresponds to a unique signal direction. Therefore, the DOA can be readily determined by measuring the signal strength at each antenna in the receiver array.
\item We formulate an optimization problem aimed at minimizing the Frobenius norm of the error between the ideal 2D DFT matrix and the EM response of the SIM, subject to the constraint that each meta-atom has a constant transmission level. Due to the non-convex constraint and cascaded multiplications of phase shifts, finding the optimal phase shift solution is non-trivial. To tackle this issue, we develop a gradient descent algorithm for iteratively updating the SIM's phase shifts to realize the desired 2D DFT function.
\item The outputs from the SIM can provide a coarse on-grid estimate of the DOA for a small number of receiver probes. To further improve the DOA estimation accuracy, we adjust the phase shifts in the zeroth layer of the SIM for each snapshot to generate a set of 2D DFT matrices having mutually orthogonal spatial frequency bins. This allows the SIM to focus the energy of the incident wave onto the specific grid point perfectly matching its direction, yielding the strongest magnitude at the matched point. The number of snapshots determines the trade-off between the estimation accuracy versus the observation time.
\item We theoretically evaluate the performance of the proposed DOA estimator by deriving an upper bound for its mean square error (MSE). As the receiver array directly observes the angular spectrum of the incident signal, the proposed DOA estimator differs fundamentally from the existing techniques relying on phase-sensitive receivers and array signal processing. Furthermore, using low-complexity energy detectors significantly reduces the hardware costs without compromising the DOA estimation accuracy.
\item Numerical results demonstrate the effectiveness of SIM to perform DOA estimation. Extensive experiments are conducted to determine the optimal SIM setups for 2D DFT for both $\left ( 2,2 \right )$ and $\left ( 4,4 \right )$ grid points. We also verify the convergence behavior of the proposed gradient descent algorithm and corroborate the accuracy of our analytical results. Specifically, the SIM using the advanced wave-based computation conceived is capable of estimating the DOA with an MSE of $10^{-4}$.
\end{enumerate}

The rest of the paper is structured as follows. Section \ref{sec2} introduces the system model of SIM-based DOA estimation. Section \ref{sec3} formulates our optimization problem and presents the gradient descent algorithm designed for optimizing the SIM to realize 2D DFT. Furthermore, Section \ref{sec4} introduces the practical estimation protocol as well as the specific procedures of DOA estimation using the SIM. Section \ref{sec5} analyzes the theoretical performance of the SIM-based DOA estimator. Additionally, Section \ref{sec6} provides simulation results to verify our analysis and evaluate the performance of the proposed estimator. Finally, Section \ref{sec7} concludes the paper and discusses potential future directions.

\emph{Notation:} Scalars are denoted by italic letters; Vectors and matrices are denoted by boldface lowercase and uppercase letters, respectively; $\Re \left \{ z \right \}$, $\Im \left \{ z \right \}$, and $\left | z \right |$ represent the real part, imaginary part, and modulus of a complex number $z$, respectively; For a complex-valued vector $\boldsymbol{v}$, $\left \| \boldsymbol{v} \right \|$ denotes its Euclidean norm, and $\textrm{diag}\left ( \boldsymbol{v} \right )$ is a diagonal matrix with the elements of $\boldsymbol{v}$ along the diagonal; For any general matrix $\boldsymbol{M}$, $\boldsymbol{M}^{\ast }$, $\boldsymbol{M}^{T}$, $\boldsymbol{M}^{H}$, $\left \| \boldsymbol{M} \right \|_{F}$, $\textrm{rank}\left ( \boldsymbol{M} \right )$, and $\left [ \boldsymbol{M} \right ]_{i,j}$ denote its conjugate, transpose, Hermitian transpose, Frobenius norm, rank, and the $\left ( i,j \right )$-th element, respectively; $\boldsymbol{M}\otimes\boldsymbol{N}$ represents the Kronecker product of the matrices $\boldsymbol{M}$ and $\boldsymbol{N}$, while $\textrm{vec}\left ( \cdot \right )$ represents the vectorization operator; $\boldsymbol{I}_{M}$ denotes an identity matrix of size $M$; $\boldsymbol{0}$ represents an all-zero matrix of appropriate dimensions; $\left \lceil x \right \rceil$ refers to the nearest integer greater than or equal to $x$; Moreover, $\mathbb{E}\left \{ \cdot \right \}$ represents the expectation operator; $\arcsin\left ( \cdot \right )$ and $\arctan\left ( \cdot \right )$ are the four-quadrant inverse sine and inverse tangent functions, respectively; $j$ is the imaginary unit; $\mathbb{C}^{x\times y}$ represents the space of $x\times y$ complex-valued matrices; $\nabla _{\boldsymbol{x}}f\left ( \boldsymbol{x} \right )$ denotes the gradient of the function $f$ with respect to (\emph{w.r.t.}) the vector $\boldsymbol{x}$; $\partial f/\partial x$ represents the partial derivative of a function $f$ \emph{w.r.t.} the variable $x$. The distribution of a circularly symmetric complex Gaussian (CSCG) random vector with mean vector $\boldsymbol{v}$ and covariance matrix $\boldsymbol{\Sigma }$ is denoted by $\sim \mathcal{CN}\left ( \boldsymbol{v}, \boldsymbol{\Sigma } \right )$, where $\sim $ stands for “is distributed as”.
\section{SIM-Based Array System Model}\label{sec2}
In this section, we present the system model for the SIM-based array used for performing DOA estimation.
\subsection{Incident Signal Model}
\begin{figure*}[!t]
\centering
\includegraphics[width=16cm]{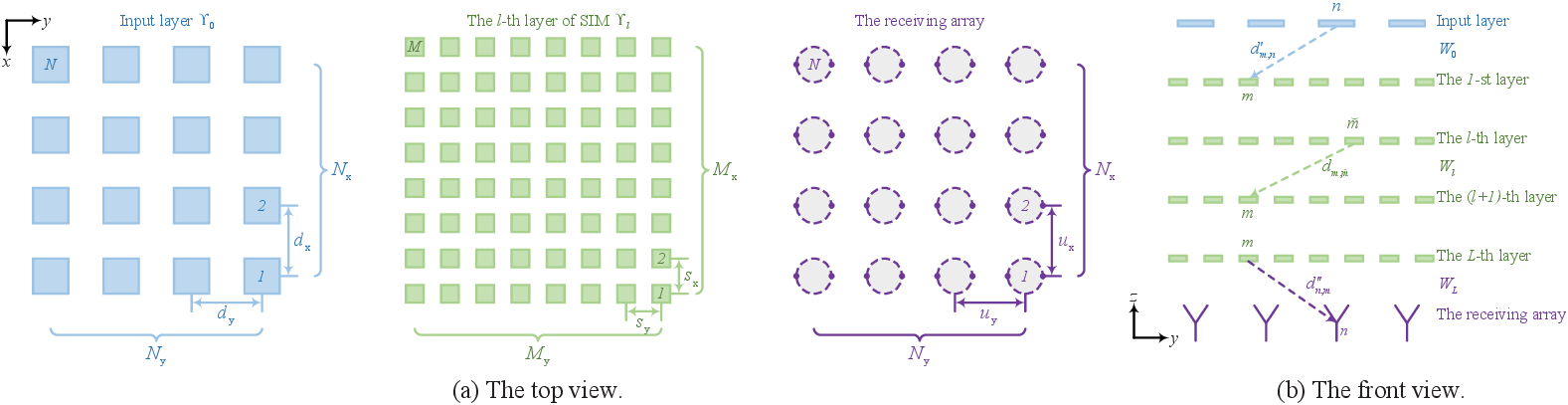}
\caption{The top and front views of the SIM-based array systems.}\vspace{-0.6cm}
\label{fig_3}
\end{figure*}
As depicted in Fig. \ref{fig_2}, we utilize a UPA placed on the ground (i.e., the $x$-$y$ plane) to estimate the DOA parameters. In contrast to conventional array-aided systems, a SIM consisting of $\left ( L+1 \right )$ metasurface layers is integrated with the UPA to transform the incident signal into its angular spectrum. We assume that the SIM is positioned horizontally, with all metasurface layers parallel to the $x$-$y$ plane. To avoid ambiguity, the metasurfaces are labeled by $0 \sim L$ from the top to the bottom, as shown in Fig. \ref{fig_3}(b). Let $\varphi \in \left [ 0,2\pi \right )$ and $\vartheta \in \left [ 0,\pi/2 \right ]$ represent the physical azimuth angle and elevation angle of the DOA of the radiation source relative to the zeroth layer of the SIM, which has $N = N_{\textrm{x}} N_{\textrm{y}}$ meta-atoms, with $N_{\textrm{x}}$ and $N_{\textrm{y}}$ representing the number of meta-atoms in the $x$- and $y$-directions, respectively. Additionally, the corresponding element spacings are $d_{\textrm{x}}$ and $d_{\textrm{y}}$. Therefore, the electrical angles $\psi _{\textrm{x}}$ and $\psi _{\textrm{y}}$ in the $x$- and $y$-directions are given by \cite{TSP_2012_Heidenreich_Joint}
\begin{align}
 \psi _{\textrm{x}}&=\kappa d_{\textrm{x}}\sin\left ( \vartheta \right )\cos\left ( \varphi \right ), \label{eq1}\\
\psi _{\textrm{y}}&=\kappa d_{\textrm{y}}\sin\left ( \vartheta \right )\sin\left ( \varphi \right ), \label{eq2}
\end{align}
respectively, where $\kappa =2\pi /\lambda $ represents the wavenumber, with $\lambda$ being the wavelength.

Hence, the steering vector \emph{w.r.t.} the zeroth layer of SIM $\boldsymbol{a}\left ( \psi _{\textrm{x}}, \psi _{\textrm{y}} \right )\in \mathbb{C}^{N \times 1}$ is written as
\begin{align}
\boldsymbol{a}\left ( \psi _{\textrm{x}}, \psi _{\textrm{y}} \right )=\boldsymbol{a}_{\textrm{y}}\left ( \psi _{\textrm{y}} \right ) \otimes \boldsymbol{a}_{\textrm{x}}\left ( \psi _{\textrm{x}} \right ), \label{eq3}
\end{align}
and the elements of the vectors $\boldsymbol{a}_{\textrm{x}}\left ( \psi _{\textrm{x}} \right )\in \mathbb{C}^{N_{\textrm{x}}\times 1}$ and $\boldsymbol{a}_{\textrm{y}}\left ( \psi _{\textrm{y}} \right )\in \mathbb{C}^{N_{\textrm{y}}\times 1}$ are defined as follows:
\begin{align}
\left [ \boldsymbol{a}_{\textrm{x}}\left ( \psi _{\textrm{x}} \right ) \right ]_{n_{\textrm{x}}}&\triangleq e^{j\psi _{\textrm{x}}\left ( n_{\textrm{x}}-1 \right )},\quad n_{\textrm{x}}=1,\cdots ,N_{\textrm{x}}, \label{eq4}\\
 \left [ \boldsymbol{a}_{\textrm{y}}\left ( \psi _{\textrm{y}} \right ) \right ]_{n_{\textrm{y}}}&\triangleq e^{j\psi _{\textrm{y}}\left ( n_{\textrm{y}}-1 \right )},\quad n_{\textrm{y}}=1,\cdots ,N_{\textrm{y}}. \label{eq5}
\end{align}

Let $s\in \mathbb{C}$ represent the signal transmitted from the radiation source\footnote{Since this is the first attempt in this area, we are considering the case of a single source for the sake of brevity.}, which is modeled as a CSCG random variable with zero mean and unit variance. Hence, the signal $\boldsymbol{x}\in \mathbb{C}^{N\times 1}$ incident upon the zeroth layer of the SIM can be expressed as
\begin{align}
 \boldsymbol{x}=\boldsymbol{a}\left ( \psi _{\textrm{x}}, \psi _{\textrm{y}} \right )s. \label{eq6}
\end{align}
\subsection{SIM Model}

The middle of Fig. 1 shows a schematic diagram of a SIM device. For the sake of conceptual simplicity, we assume that the $L$ intermediate metasurface layers are each modeled as a UPA having isomorphic arrangements. Additionally, we assume that the $\left ( L+1 \right )$ metasurfaces are evenly spaced. Let $T_{\textrm{SIM}}$ represent the thickness of the SIM. As such, the vertical spacing between the adjacent layers is obtained by $s_{\textrm{layer}}=T_{\textrm{SIM}}/L$. In practice, the SIM is enclosed in a supporting structure surrounded by wave-absorbing materials, to prevent interferences from undesired diffraction, scattering, and environmental noise \cite{NE_2022_Liu_A}. As shown in Fig. \ref{fig_3}, each metasurface layer consists of $M = M_{\textrm{x}} M_{\textrm{y}}$ meta-atoms, where $M_{\textrm{x}}$ and $M_{\textrm{y}}$ are the number of meta-atoms in the $x$- and $y$-directions, respectively. Moreover, the corresponding spacings between the adjacent meta-atoms on the intermediate layers are set to $s_{\textrm{x}}$ and $s_{\textrm{y}}$.

As stated earlier, each meta-atom is capable of adjusting the phase shift of the EM waves passing through it by controlling the bias voltage of the associated circuit \cite{JSAC_2023_An_Stacked, arXiv_2023_An_Stacked, NE_2022_Liu_A}. Let $\boldsymbol{\upsilon}_{l}=\left [\upsilon_{l,1},\upsilon_{l,2},\cdots ,\upsilon_{l,M} \right ]^{T}\in \mathbb{C}^{M\times 1},\ l = 1,\cdots ,L$ represent the complex-valued transmission coefficient vector for the $l$-th layer, where $\upsilon_{l,m}=e^{j\xi _{l,m}},\ m=1, \cdots, M,\ l=1, \cdots, L$ with $ \xi _{l,m}\in \left [ 0,2\pi \right )$ representing the phase shift of the $m$-th meta-atom on the $l$-th layer \cite{JSAC_2023_An_Stacked}. Furthermore, let $\boldsymbol{\Upsilon}_{l} =\textrm{diag}\left ( \boldsymbol{\upsilon }_{l} \right )\in \mathbb{C}^{M\times M}$ represent the corresponding transmission coefficient matrix for the $l$-th layer. In particular, let $\boldsymbol{\upsilon}_{0}=\left [\upsilon_{0,1},\upsilon_{0,2},\cdots ,\upsilon_{0,N} \right ]^{T}\in \mathbb{C}^{N\times 1}$ and $\boldsymbol{\Upsilon}_{0} =\textrm{diag}\left ( \boldsymbol{\upsilon }_{0} \right )\in \mathbb{C}^{N\times N}$ denote the complex-valued transmission coefficient vector and the corresponding matrix for the input layer (i.e., the zeroth layer in Fig. 1), respectively, where we have $\upsilon_{0,n}=e^{j\xi _{0,n}},\ n=1, \cdots, N$ and $\xi _{0,n}$ denotes the phase shift of the $n$-th meta-atom on the zeroth layer.

Furthermore, let $\boldsymbol{W}_{l}\in \mathbb{C}^{M\times M},\ l = 1,\cdots ,L-1$ characterize the EM wave propagation between the adjacent layers in the SIM. Specifically, the element at the $m$-th row and $\breve{m}$-th column of $\boldsymbol{W}_{l}$ represents the attenuation coefficient between the $\breve{m}$-th meta-atom on layer $l$ and the $m$-th meta-atom on layer $\left ( l+1 \right )$. Based on the Rayleigh-Sommerfeld diffraction equation \cite{BOOK_2005_Goodman_Introduction, JSAC_2023_An_Stacked}, $\left [ \boldsymbol{W}_{l} \right ]_{m,\breve{m}}$ is determined as follows:
\begin{align}
\left [ \boldsymbol{W}_{l} \right ]_{m,\breve{m}}=\frac{A_{\textrm{meta-atom}}s_{\textrm{layer}}}{2\pi d_{m,\breve{m}}^{3}}\left ( 1-j\kappa d_{m,\breve{m}}\right )e^{j \kappa d_{m,\breve{m}}}, \label{eq7}
\end{align}
where $A_{\textrm{meta-atom}}$ denotes the area of each meta-atom, and $d_{m,\breve{m}}$ represents the corresponding propagation distance, which is calculated as follows:
\begin{align}
 d_{m,\breve{m}}=\sqrt{\left ( m_{\textrm{x}}-\breve{m}_{\textrm{x}} \right )^{2}s_{\textrm{x}}^{2}+\left ( m_{\textrm{y}}-\breve{m}_{\textrm{y}} \right )^{2}s_{\textrm{y}}^{2}+s_{\textrm{layer}}^{2}}, \label{eq8}
\end{align}
with $m_{\textrm{y}}$ and $m_{\textrm{x}}$ being defined by
\begin{align}
 m_{\textrm{y}}&\triangleq \left \lceil m/M_{\textrm{x}} \right \rceil, \label{eq9}\\
 m_{\textrm{x}}&\triangleq m-\left ( m_{\textrm{y}}-1 \right )M_{\textrm{x}}, \label{eq10}
\end{align}
Similarly, $\breve{m}_{\textrm{y}}$ and $\breve{m}_{\textrm{x}}$ are obtained by replacing $m$ in \eqref{eq9} and \eqref{eq10} with $\breve{m}$.

Furthermore, let $\boldsymbol{W}_{0}\in \mathbb{C}^{M\times N}$ represent the attenuation coefficient matrix between the input layer and the first layer. The $\left ( m,n \right )$-th entry of $\boldsymbol{W}_{0}$, denoted by $\left [ \boldsymbol{W}_{0} \right ]_{m,n}$, is obtained by replacing $d_{m,\breve{m}}$ in \eqref{eq7} with the corresponding propagation distance ${d}'_{m,n}$, which is calculated using \eqref{eq11}, as shown at the top of this page, where we have
\begin{figure*}
\begin{small}\begin{align}
 {d}'_{m,n}=\sqrt{\left [ \left ( m_{\textrm{x}}-\frac{1+M_{\textrm{x}}}{2} \right )s_{\textrm{x}}-\left ( n_{\textrm{x}}-\frac{1+N_{\textrm{x}}}{2} \right )d_{\textrm{x}} \right ]^{2}+\left [ \left ( m_{\textrm{y}}-\frac{1+M_{\textrm{y}}}{2} \right )s_{\textrm{y}}-\left ( n_{\textrm{y}}-\frac{1+N_{\textrm{y}}}{2} \right )d_{\textrm{y}} \right ]^{2}+s_{\textrm{layer}}^{2}}, \label{eq11}
\end{align}\end{small}\vspace{-0.6cm}
\hdashrule[0.5ex][c]{16cm}{1pt}{3mm}
\end{figure*}
\begin{align}
 n_{\textrm{y}}&\triangleq \left \lceil n/N_{\textrm{x}} \right \rceil, \label{eq12}\\
 n_{\textrm{x}}&\triangleq n-\left ( n_{\textrm{y}}-1 \right )N_{\textrm{x}}. \label{eq13}
\end{align}

Similarly, let $\boldsymbol{W}_{L}\in \mathbb{C}^{N\times M}$ represent the attenuation coefficient matrix between the $L$-th metasurface layer and the output layer, i.e., the receiver antenna array. The receiver is a UPA arranged in the same pattern as the zeroth layer of SIM, and it is placed at $s_{\textrm{layer}}$ meters away from layer $L$. It is clear that $\boldsymbol{W}_{L}$ and $\boldsymbol{W}_{0}$ exhibit symmetry. Thus, we have $\boldsymbol{W}_{L} = \boldsymbol{W}_{0}^{T}$.

As a result, the overall forward propagation process through the SIM $\boldsymbol{G}\in \mathbb{C}^{N\times N}$ is described as
\begin{align}
\boldsymbol{G}=\boldsymbol{W}_{L}\boldsymbol{\Upsilon} _{L}\boldsymbol{W}_{L-1}\cdots \boldsymbol{W}_{2}\boldsymbol{\Upsilon} _{2}\boldsymbol{W}_{1}\boldsymbol{\Upsilon} _{1}\boldsymbol{W}_{0}. \label{eq15}
\end{align}
\subsection{Received Signal Model}
As mentioned earlier, the receiver is a UPA consisting of $N = N_{\textrm{x}} N_{\textrm{y}}$ receiver antennas. For a single source transmitting a waveform $s$, the complex signal vector $\boldsymbol{r}\in \mathbb{C}^{N\times 1}$ received at the array can be expressed as
\begin{align}
\boldsymbol{r}=\sqrt{\varrho }\boldsymbol{G}\boldsymbol{\Upsilon} _{0}\boldsymbol{x}+\boldsymbol{u}=\sqrt{\varrho }\boldsymbol{G}\boldsymbol{\Upsilon} _{0}\boldsymbol{a}\left ( \psi _{\textrm{x}},\psi _{\textrm{y}} \right )s+\boldsymbol{u}, \label{eq16}
\end{align}
where $\varrho$ denotes the SNR, and $\boldsymbol{u}\in \mathbb{C}^{N\times 1}$ is the measurement noise vector at the receiver array, which is modeled as a CSCG random vector satisfying $\boldsymbol{u}\sim \mathcal{CN}\left ( \boldsymbol{0}, \boldsymbol{I}_{N} \right )$. It is also assumed that $s$ and $\boldsymbol{u}$ are uncorrelated. 
\begin{remark}
In contrast to the conventional array-aided system, the received signal in \eqref{eq16} has undergone a controllable analog transformation using the SIM. By appropriately configuring the SIM's phase shifts, the receiver antenna array may directly observe the angular spectrum of the incident signal. This would substantially simplify both the hardware design and the subsequent signal processing, while also reducing the energy consumption.
\end{remark}
\begin{remark}
Although $\boldsymbol{G}$ in \eqref{eq15} involves a large number of matrix multiplications, it is important to note that these operations occur automatically at the speed of light as the incoming signal passes through each layer of the SIM. This advanced wave-based computing paradigm enables the calculations to be completed in nanoseconds. Additionally, the inherent parallel processing capability of the SIM means that the calculation time is independent of the number of meta-atoms per layer.
\end{remark}
\section{SIM Optimization for Realizing 2D DFT}\label{sec3}
To achieve the expected DOA estimation capability, we need to guide the SIM to output the angular spectrum. In this section, we formulate an optimization problem for implementing the 2D DFT in the wave domain and devise a gradient descent algorithm to find a high-quality near-optimal solution for the phase shifts.

\subsection{Optimization Problem}
Specifically, for the 2D DFT matrix $\boldsymbol{F}\in \mathbb{C}^{N\times N}$ of $N = N_{\textrm{x}}N_{\textrm{y}}$ grid points, its $\left ( n,\breve{n} \right )$-th entry is defined as follows:
\begin{align}
 f_{n,\breve{n}} = \left [ \boldsymbol{F} \right ]_{n,\breve{n}}\triangleq e^{-j2\pi \frac{\left ( n_{\textrm{x}}-1 \right )\left ( \breve{n}_{\textrm{x}}-1 \right )}{N_{\textrm{x}}}}e^{-j2\pi \frac{\left ( n_{\textrm{y}}-1 \right )\left ( \breve{n}_{\textrm{y}}-1 \right )}{N_{\textrm{y}}}}, \label{eq17}
\end{align}
where $n_{\textrm{y}}$ and $n_{\textrm{x}}$ are as defined in \eqref{eq12} and \eqref{eq13}, while $\breve{n}_{\textrm{y}}$ and $\breve{n}_{\textrm{x}}$ are obtained upon replacing $n$ by $\breve{n}$.

To evaluate the accuracy of the SIM's response fitting the 2D DFT matrix, we employ the Frobenius norm of the fitting error between each target entry and the EM response of the SIM. Specifically, the loss function $\mathcal{L}$ is defined as
\begin{align}
 \mathcal{L} = \left \| \beta \boldsymbol{G}-\boldsymbol{F} \right \|_{F}^{2}, \label{eq18}
\end{align}
where $\beta \in \mathbb{C}$ represents the scaling factor used for keeping the SIM's response at the required normalized value.

\begin{remark}
Note that in \eqref{eq18} we are not actually multiplying the received signal by an extra coefficient $\beta$. Multiplying $\boldsymbol{G}$ by a scaling factor is only used for ensuring that the error is normalized to the same level for a fair comparison.
\end{remark}

To minimize the loss function in \eqref{eq18}, the optimization problem of matching the SIM's response to the 2D DFT matrix is formulated as
\begin{subequations}\label{eq19}
\begin{alignat}{2}
&\!\min_{\left \{ \xi _{l,m} \right \}} &\quad & \mathcal{L} = \left \| \beta \boldsymbol{G}-\boldsymbol{F} \right \|_{F}^{2} \label{eq19a}\\
&\textrm{s.t.} & & \boldsymbol{G}=\boldsymbol{W}_{L}\boldsymbol{\Upsilon} _{L}\boldsymbol{W}_{L-1}\cdots \boldsymbol{W}_{2}\boldsymbol{\Upsilon} _{2}\boldsymbol{W}_{1}\boldsymbol{\Upsilon} _{1}\boldsymbol{W}_{0}, \label{eq19b}\\
& & & \boldsymbol{\Upsilon}_{l}=\textrm{diag}\left ( \left [e^{j\xi _{l,1}},e^{j\xi _{l,2}},\cdots ,e^{j\xi _{l,M}} \right ]^{T}\right ), \label{eq19c}\\
& & & \xi _{l,m} \in \left [ 0,2\pi \right ),\, m=1, \cdots, M,\, l=1, \cdots, L, \label{eq19d}\\
& & & \beta \in \mathbb{C}. \label{eq19e}
\end{alignat}
\end{subequations}

Note that due to the non-convex constant modulus constraint and the fact that the phase shifts associated with different metasurface layers are highly coupled, the problem in \eqref{eq19} is non-trivial to solve optimally. In the subsequent subsection, we customize a gradient descent method for efficiently finding a near-optimal solution to \eqref{eq19}.
\subsection{Proposed Gradient Descent Algorithm}\label{sec3_2}
The popular gradient descent algorithm iteratively adjusts the phase shifts of the SIM for gradually minimizing the loss function in \eqref{eq19a}. Specifically, gradient descent involves two main procedures: \emph{i)} calculating the derivative; and \emph{ii)} updating the parameters.
\subsubsection{Derivative Calculation}
For a tentative SIM model, the gradient of the loss function $\mathcal{L}$ \emph{w.r.t.} the phase shift vector $\boldsymbol{\xi} _{l}$ of the $l$-th layer in a SIM is calculated by
\begin{align}
\nabla_{\boldsymbol{\xi} _{l}} \mathcal{L} &=2\sum_{n=1}^{N}\Im\left \{ \beta ^{\ast } \boldsymbol{\Upsilon} _{l}^{H}\boldsymbol{P}_{l,n}^{H}\left ( \beta \boldsymbol{g}_{n}-\boldsymbol{f}_{n} \right ) \right \}, \label{eq20}
\end{align}
for $l=1, \cdots, L$, where $\boldsymbol{g}_{n}\in \mathbb{C}^{N\times 1},\ n=1, \cdots, N$ and $\boldsymbol{f}_{n}\in \mathbb{C}^{N\times 1},\ n=1, \cdots, N$ represent the $n$-th column of $\boldsymbol{G}$ and $\boldsymbol{F}$, respectively. Furthermore, $\boldsymbol{P}_{l,n}\in \mathbb{C}^{N\times M},\ n=1, \cdots, N,\ l=1, \cdots, L$ denotes the equivalent coefficient matrix associated with the $l$-th metasurface layer activating the $n$-th meta-atom on the input layer, satisfying $\boldsymbol{P}_{l,n}\boldsymbol{\upsilon }_{l}=\boldsymbol{g}_{n}$. Specifically, $\boldsymbol{P}_{l,n}$ is defined as
\begin{align}
 \boldsymbol{P}_{l,n}= \boldsymbol{W}_{L}\boldsymbol{\Upsilon} _{L}\boldsymbol{W}_{L-1}\cdots \boldsymbol{W}_{l+1}\boldsymbol{\Upsilon} _{l+1}\boldsymbol{W}_{l}\textrm{diag}\left ( \boldsymbol{q}_{l,n} \right ),\label{eq21}
\end{align}
with $\boldsymbol{q}_{l,n}\in \mathbb{C}^{M\times 1}$ representing the complex signal component illuminating the $l$-th layer of the SIM from the $n$-th meta-atom in the zeroth layer, defined as
 \begin{align}
\boldsymbol{q}_{l,n}&=\boldsymbol{W}_{l-1}\boldsymbol{\Upsilon} _{l-1}\boldsymbol{W}_{l-2}\cdots \boldsymbol{W}_{2}\boldsymbol{\Upsilon} _{2}\boldsymbol{W}_{1}\boldsymbol{\Upsilon} _{1}\boldsymbol{w}_{0,n},
\end{align}
for $n=1, \cdots, N,\ l = 1,\cdots ,L$, where $\boldsymbol{w}_{0,n}\in \mathbb{C}^{M\times 1}$ represents the $n$-th column of $\boldsymbol{W}_{0}$. Please refer to Appendix \ref{A1} for the detailed procedures showing the derivation of \eqref{eq20}.
\begin{remark}
Note that the derivative of the cost function \emph{w.r.t.} each layer in \eqref{eq20} depends on an intermediate variable that is a product of the phase shift matrix and the attenuation coefficient matrix, starting from the final layer and moving backward to the current layer. By storing and recursively updating this intermediate variable, we can prevent redundant calculations of the multiplications and efficiently determine the derivatives for all layers.
\end{remark}

\subsubsection{Parameter Update}
Once all the gradients \emph{w.r.t.} the SIM's phase shift vectors have been calculated, we simultaneously update the phase shift values $\boldsymbol{\xi} _{l}$ in the specific direction that decreases the loss function value. At each iteration, the update formula is as follows:
\begin{align}
\boldsymbol{\xi }_{l}\leftarrow \boldsymbol{\xi }_{l}-\eta \nabla_{\boldsymbol{\xi }_{l}}\mathcal{L}, \label{eq23}
\end{align}
where $\eta >0$ represents the learning rate. To ensure a stable convergence, the learning rate $\eta$ also decreases over time. Specifically, we have
\begin{align}
 \eta\leftarrow \eta \zeta , \label{eq24}
\end{align}
with $\zeta $ representing the decay parameter.

Additionally, the auxiliary scaling factor $\beta$ also has to be updated during each iteration to maintain the required normalized level. Specifically, given a tentative SIM response matrix $\boldsymbol{G}$, the optimal value of $\beta$ can be readily obtained by utilizing the least squares method, yielding
\begin{align}
 \beta = \left ( \boldsymbol{g} ^{H}\boldsymbol{g} \right )^{-1} \boldsymbol{g}^{H}\boldsymbol{f}, \label{eq25_0}
\end{align}
where we have $\boldsymbol{g} = \textrm{vec}\left ( \boldsymbol{G} \right )$ and $\boldsymbol{f} = \textrm{vec}\left ( \boldsymbol{F} \right )$.

The phase shift values are updated repeatedly until either the loss function $\mathcal{L}$ converges or the number of iterations achieves the maximum tolerable value. To summarize briefly, the general procedure of the proposed gradient descent algorithm is outlined in Table \ref{alg1}.
\begin{algorithm}[t]
\caption{The Gradient Descent Algorithm Proposed for Optimizing the SIM's Phase Shifts.}
\label{alg1}
\begin{algorithmic}[1]
\STATE \textbf{Input:} $\boldsymbol{F}$, $\boldsymbol{W} _{l},\, l=0,1,\cdots ,L$.
\STATE Randomly initialize all phase shift values $\boldsymbol{\xi} _{l},\, l=1,\cdots ,L$ by sampling from a uniform distribution.
\STATE Calculate the current loss function $\mathcal{L}$.
\STATE \textbf{Repeat}
\STATE \hspace{0.5cm} {Calculate the derivatives for each layer using \eqref{eq20}.}
\STATE \hspace{0.5cm} Update the phase shift vectors $\boldsymbol{\xi} _{l},\, l=1,\cdots ,L$ using \eqref{eq23};
\STATE \hspace{0.5cm} Adjust the learning rate $\eta$ using \eqref{eq24};
\STATE \hspace{0.5cm} Adjust the scaling factor $\beta$ using \eqref{eq25_0};
\STATE \textbf{Until} the fractional reduction in $\mathcal{L} $ falls below a preset threshold or the maximum number of iterations is reached.
\STATE \textbf{Output:} $\left \{ \boldsymbol{\xi} _{1},\boldsymbol{\xi} _{2},\cdots ,\boldsymbol{\xi} _{L} \right \}$.
\end{algorithmic}
\end{algorithm}
\subsection{The Choice for $M$}
In this subsection, we briefly discuss the SIM parameter selection for achieving the desired 2D DFT functionality.
\begin{theorem}\label{theorem1}
 A necessary condition for achieving $\mathcal{L}=0$ is that $M\geq N$.
\end{theorem}

\emph{Proof:} If $M<N$, the rank of $\boldsymbol{G}$ will be limited to $\textrm{rank}\left ( \boldsymbol{G} \right )\leq \textrm{rank}\left ( \boldsymbol{W}_{0} \right )\leq \min\left ( N,M \right ) = M <N$. As a result, the SIM's response matrix $\boldsymbol{G}$ having arbitrary phase shift values can no longer accurately fit the 2D DFT matrix $\boldsymbol{F}$, whose rank is $\textrm{rank}\left ( \boldsymbol{F} \right )=N$. This proves the theorem. $\hfill \square$

\begin{remark}
By employing a SIM to implement the 2D DFT in the wave domain, the system can directly generate the angular spectrum at the receiver array and provide an on-grid estimate of the DOA parameters for the incident signal. However, this may result in a coarse estimate with limited precision for a small value of $N$. To achieve high DOA estimation accuracy, this requires employing a large number of probes at the receiver, which is not practical due to both the physical space and cost limitations. Additionally, accurately implementing the 2D DFT associated with arbitrary grid points requires a huge number $M$ of meta-atoms on each layer. While the unit price is reasonable, the entire SIM would be costly. Fortunately, the zeroth layer provides an extra design degree of freedom (DoF), which can be exploited for generating a set of 2D DFT matrices associated with different frequency bins. This has the potential of substantially improving the DOA estimation accuracy of a moderate-size SIM, as it will be discussed further in Section \ref{sec4}.
\end{remark}

\section{SIM-Based DOA Estimation}\label{sec4}
\begin{figure*}[!t]
\centering
\includegraphics[width=16cm]{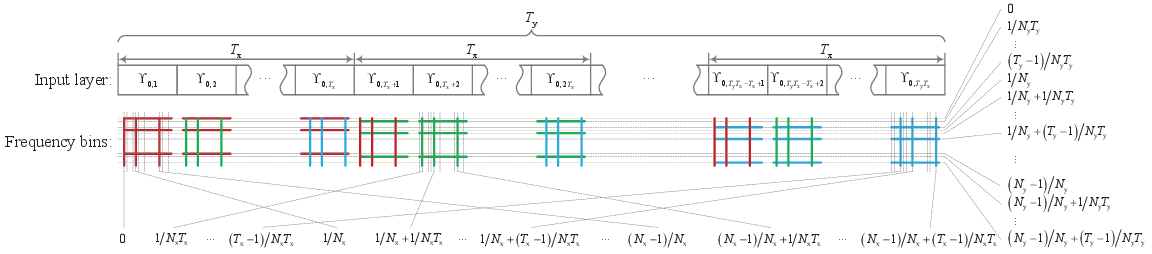}
\caption{The proposed SIM-based DOA estimation protocol.}\vspace{-0.6cm}
\label{fig_4}
\end{figure*}
In this section, we first introduce the proposed SIM-based DOA estimation protocol by appropriately configuring the phase shift values of the zeroth layer, i.e., $\boldsymbol{\Upsilon} _{0}$. We then present the specific DOA estimation procedure based on this configuration.
\subsection{Estimation Protocol}\label{sec4_1}
As shown in Fig. 3, the proposed protocol divides the total observation time $T$ into $T_{\textrm{y}}$ blocks, each of length $T_{\textrm{x}}$, so that $T=T_{\textrm{x}}T_{\textrm{y}}$. The phase shift vectors for the first to the $L$-th layers are determined by employing the optimization process described in Section \ref{sec3_2} and remain the same during $T$ snapshots. By contrast, the phase shift vector for the zeroth layer of the SIM is reconfigured at each time slot in order to generate a set of DFT matrices having orthogonal spatial frequency bins. Specifically, at the $t$-th snapshot, the phase shift of the $n$-th meta-atom on the zeroth layer is configured as
\begin{align}
 \xi _{0,n,t} =-2\pi \frac{\left ( n_{\textrm{x}}-1 \right )\left ( t_{\textrm{x}}-1 \right )}{N_{\textrm{x}}T_{\textrm{x}}}-2\pi \frac{\left ( n_{\textrm{y}}-1 \right )\left ( t_{\textrm{y}}-1 \right )}{N_{\textrm{y}}T_{\textrm{y}}}, \label{eq25}
\end{align}
where $n_{\textrm{y}}$ and $n_{\textrm{x}}$ are defined by \eqref{eq12} and \eqref{eq13}, $t_{\textrm{y}}$ and $t_{\textrm{x}}$ represent the block index and the time slot index within that block, respectively, which are defined by
\begin{align}
 t_{\textrm{y}}&\triangleq \left \lceil t/T_{\textrm{x}} \right \rceil, \label{eq26}\\
 t_{\textrm{x}}&\triangleq t-\left ( t_{\textrm{y}}-1 \right )T_{\textrm{x}}. \label{eq27}
\end{align}

Note that upon right-multiplying $\boldsymbol{G}$ (i.e., the well-fitted version of $\boldsymbol{F}$) by $\boldsymbol{\Upsilon }_{0,t}$, the SIM implicitly characterizes a set of 2D DFT matrices whose frequency bins are all mutually orthogonal to each other.
\subsection{SIM-Based DOA Estimator}\label{sec4_2}
Under the noiseless received signal model, the EM waves propagating through the optimized SIM are automatically focused on the specific antenna index and snapshot index corresponding to the on-grid DOA estimate and spatial frequency offset component of the incoming signal, respectively. As a result, the DOA parameters of the incoming signal can be readily estimated by measuring the energy distribution across the receiver antenna array, which is in contrast to conventional DOA estimation algorithms relying on phase-sensitive receivers and array signal processing.

Specifically, let $r_{n,t}$ represent the signal received at the $n$-th probe in the $t$-th snapshot. After collecting the received signals over $T$ snapshots, we then search for the index of the strongest signal magnitude. The 2D index of the peak is obtained as follows:
\begin{align}
\left [ \hat{n},\hat{t} \right ] = \arg\underset{n=1,\cdots ,N,\atop t=1,\cdots , T}{\max}\left | r_{n,t} \right |^{2}. \label{eq28}
\end{align}

Therefore, the corresponding electrical angles of the incident signal are obtained by
\begin{align}
 \hat{\psi}_{\textrm{x}} &= \textrm{mod}\left [ 2\left ( \frac{\hat{n}_{\textrm{x}}-1}{N_{\textrm{x}}} +\frac{\hat{t}_{\textrm{x}}-1}{N_{\textrm{x}}T_{\textrm{x}}} \right )+1,2 \right ]-1, \label{eq29}\\
 \hat{\psi}_{\textrm{y}} &=\textrm{mod}\left [ 2\left ( \frac{\hat{n}_{\textrm{y}}-1}{N_{\textrm{y}}} +\frac{\hat{t}_{\textrm{y}}-1}{N_{\textrm{y}}T_{\textrm{y}}} \right )+1,2 \right ]-1, \label{eq30}
\end{align}
respectively, where $\hat{n}_{\textrm{y}}$ and $\hat{n}_{\textrm{x}}$ are obtained by substituting $\hat{n}$ into \eqref{eq12} and \eqref{eq13}, respectively, while $\hat{t}_{\textrm{y}}$ and $\hat{t}_{\textrm{x}}$ are obtained by substituting $\hat{t}$ into \eqref{eq26} and \eqref{eq27}, respectively.

Based on \eqref{eq29} and \eqref{eq30}, the estimated azimuth and elevation angles $\hat{\varphi }$ and $\hat{\vartheta }$ are given by
\begin{align}
 \hat{\varphi }&=\arctan\left ( \frac{\psi _{\textrm{y}}d_{\textrm{x}}}{\psi _{\textrm{x}}d_{\textrm{y}}} \right ), \label{eq31}\\
 \hat{\vartheta }&=\arcsin\left (\frac{1}{\kappa } \sqrt{\frac{\psi _{\textrm{x}}^{2}}{d_{\textrm{x}}^{2}}+\frac{\psi _{\textrm{y}}^{2}}{d_{\textrm{y}}^{2}}} \right ). \label{eq32}
\end{align}

\begin{remark}
The parameter $T$ strikes a flexible tradeoff between the estimation accuracy and the number of snapshots needed. Increasing the number of snapshots can enhance the estimation accuracy, but this requires increasing the switching speed of the FPGA controlling the zeroth layer in order to collect more observations within a given period.
\end{remark}
\begin{remark}
In conventional radar systems, the antenna arrays first receive signals and down-convert them to baseband signals before estimating the DOAs. Again, the SIM operates in a fundamentally different way by directly processing the received RF signals, without the need for an individual RF chain and ADC at each antenna element. This substantially mitigates both the hardware cost and energy consumption, which has great potential for onboard applications such as employing a SIM on a drone to probe the DOA of ground targets.
\end{remark}

\section{Performance Analysis}\label{sec5}
Due to the measurement noise at the receiver array, the peak index may be incorrectly identified, leading to estimation error. In this section, we analyze the performance of the proposed SIM-based DOA estimator by theoretically deriving the upper bound for its MSE\footnote{For notational brevity, we evaluate the MSE of the electrical angles instead of the azimuth and elevation values. The MSE of the azimuth and elevation angles can be readily obtained by substituting those estimated values according to \eqref{eq31} and \eqref{eq32} into \eqref{eq36} and \eqref{eq37}.}. Specifically, let $\bar{\psi}_{\text{x}}$ and $\bar{\psi}_{\text{y}}$ represent the true electrical angles of the incident signal. Hence, the MSE of the SIM-based DOA estimator is calculated by
\begin{align}
 \textrm{MSE} _{\psi_{\text{x}}}&=\sum_{n=1}^{N}\sum_{t=1}^{T}\left ( \bar{\psi}_{\text{x}}-\hat{\psi} _{\text{x},n,t} \right )^{2}\textrm{Pr}\left ( n,t \right ), \label{eq33}\\
 \textrm{MSE} _{\psi_{\text{y}}}&=\sum_{n=1}^{N}\sum_{t=1}^{T}\left ( \bar{\psi}_{\text{y}}-\hat{\psi} _{\text{y},n,t} \right )^{2}\textrm{Pr}\left ( n,t \right ), \label{eq34}
\end{align}
where $\hat{\psi} _{\text{x},n,t}$ and $\hat{\psi} _{\text{y},n,t}$ represent the estimated values of $\psi_{\text{x}}$ and $\psi_{\text{y}}$, respectively, for the case when $r_{n,t}$ has the highest energy, and $\textrm{Pr}\left ( n,t \right )$ represents the probability of detecting $r_{n,t}$ having the highest magnitude, which is defined as
\begin{align}
 \textrm{Pr}\left ( n,t \right )=\textrm{Pr}\left \{ \left | r_{n,t} \right |^{2}=\underset{\tilde{n}=1,\cdots ,N,\atop \tilde{t}=1,\cdots , T}{\max} \left | r_{\tilde{n},\tilde{t}} \right |^{2} \right \}.
\end{align}

Furthermore, the upper bounds for $\textrm{MSE}_{\psi_{\text{x}}}$ and $\textrm{MSE}_{\psi_{\text{y}}}$ are summarized in \emph{Theorem \ref{theorem2}}.
\begin{theorem}\label{theorem2}
 The analytical expressions for \eqref{eq33} and \eqref{eq34} are upper-bounded by:
 \begin{align}
 \textrm{MSE}_{\psi_{\text{x}}}\leq &\sum_{n=1}^{N}\sum_{t=1}^{T}\left ( \bar{\psi}_{\text{x}} - \hat{\psi}_{\text{x},n,t}\right )^{2}  \notag \\
 &\times Q\left ( \frac{9h_{n,t}\sqrt[3]{\frac{b_{n,t}}{h_{n,t}}}-9h_{n,t}+2}{2} \right ), \label{eq36}\\
 \textrm{MSE}_{\psi_{\text{y}}}\leq &\sum_{n=1}^{N}\sum_{t=1}^{T}\left ( \bar{\psi}_{\text{y}} - \hat{\psi}_{\text{y},n,t}\right )^{2} \notag \\
 &\times Q\left ( \frac{9h_{n,t}\sqrt[3]{\frac{b_{n,t}}{h_{n,t}}}-9h_{n,t}+2}{2} \right ), \label{eq37}
\end{align}
\end{theorem}
where we have
\begin{align}
h_{n,t}=&\mu_{2,n,t}^{3}/\mu_{3,n,t}^{2}, \label{eq38}\\
b_{n,t}=&h_{n,t}-\mu_{1,n,t}\sqrt{\frac{h_{n,t}}{\mu_{2,n,t}}}, \label{eq39}\\
\mu_{i,n,t}=&\left ( -1 \right )^{i}\left ( 2+i\left | \sqrt{2\varrho }\tilde{\boldsymbol{g}}_{\breve{n}}^{H} \boldsymbol{\Upsilon }_{0,\breve{t}}\boldsymbol{a}\left ( \bar{\psi }_{\textrm{x}},\bar{\psi }_{\textrm{y}} \right )s \right |^{2} \right ) \notag\\
&+2+i\left | \sqrt{2\varrho }\tilde{\boldsymbol{g}}_{n}^{H} \boldsymbol{\Upsilon }_{0,t}\boldsymbol{a}\left ( \bar{\psi }_{\textrm{x}},\bar{\psi }_{\textrm{y}} \right )s \right |^{2}, \label{eq40}
\end{align}
for $i=1,2,3$, $\tilde{\boldsymbol{g}}_{n}^{H}\in \mathbb{C}^{1\times N}$ and $\tilde{\boldsymbol{g}}_{\breve{n}}^{H}\in \mathbb{C}^{1\times N}$ represent the $n$-th and $\breve{n}$-th rows of $\boldsymbol{G}$, respectively. Moreover, $\breve{n}$ and $\breve{t}$ represent the antenna and snapshot indices associated with the highest energy under the noiseless condition, defined by
\begin{align}
\left [ \breve{n},\breve{t} \right ]=\arg \underset{n=1,\cdots ,N,\atop t=1,\cdots , T}{\max} \left | \tilde{\boldsymbol{g}}_{n}^{H} \boldsymbol{\Upsilon }_{0,t}\boldsymbol{a}\left ( \bar{\psi }_{\textrm{x}},\bar{\psi }_{\textrm{y}} \right )s \right |^{2}. \label{eq41}
\end{align}

\emph{Proof:} Please refer to Appendix \ref{A2}. $\hfill \square$

\section{Simulation Results}\label{sec6}
\begin{table*}[!t]
\centering
\scriptsize
\caption{The normalized loss function value (in dB) of utilizing a SIM to fit 2D DFT matrix with $\left ( 2,2 \right )$ grid points.}
\label{tab1}
\begin{tabular}{ccccccccccc}
\hline
\multicolumn{11}{c}{The first-round experiment with coarse granularity} \\ \hline
\multicolumn{1}{c|}{\multirow{2}{*}{$M$}} &\multicolumn{1}{c|}{\multirow{2}{*}{$T_{\textrm{SIM}}$}} & \multicolumn{3}{c|}{$s_{\textrm{x}}=s_{\textrm{y}}=2\lambda /3$} & \multicolumn{3}{c|}{$s_{\textrm{x}}=s_{\textrm{y}}=2\lambda /6$} & \multicolumn{3}{c}{$s_{\textrm{x}}=s_{\textrm{y}}=2\lambda /9$} \\ \cline{3-11}
\multicolumn{1}{c|}{} & \multicolumn{1}{c|}{} & \multicolumn{1}{c|}{$L=3$} & \multicolumn{1}{c|}{$L=6$} & \multicolumn{1}{c|}{$L=9$} & \multicolumn{1}{c|}{$L=3$} & \multicolumn{1}{c|}{$L=6$} & \multicolumn{1}{c|}{$L=9$} & \multicolumn{1}{c|}{$L=3$} & \multicolumn{1}{c|}{$L=6$} & $L=9$ \\ \hline
\multicolumn{1}{c|}{\multirow{3}{*}{$\ \,\ \,9$}} & \multicolumn{1}{c|}{$3\lambda $} & \multicolumn{1}{r|}{$-9.04$} & \multicolumn{1}{r|}{$-9.22$} & \multicolumn{1}{r|}{$-5.10$} & \multicolumn{1}{r|}{$-2.34$} & \multicolumn{1}{r|}{$-3.10$} & \multicolumn{1}{r|}{$-3.82$} & \multicolumn{1}{r|}{$-1.40$} & \multicolumn{1}{r|}{$-2.67$} & \multicolumn{1}{r}{$-1.28$} \\ \cline{2-11}
\multicolumn{1}{c|}{} & \multicolumn{1}{c|}{$6\lambda $} & \multicolumn{1}{r|}{$-3.72$} & \multicolumn{1}{r|}{$-15.39$} & \multicolumn{1}{r|}{$-10.59$} & \multicolumn{1}{r|}{$-1.33$} & \multicolumn{1}{r|}{$-1.39$} & \multicolumn{1}{r|}{$-1.75$} & \multicolumn{1}{r|}{$-1.10$} & \multicolumn{1}{r|}{$-1.25$} & \multicolumn{1}{r}{$-1.25$} \\ \cline{2-11}
\multicolumn{1}{c|}{} & \multicolumn{1}{c|}{$9\lambda $} & \multicolumn{1}{r|}{$-2.03$} & \multicolumn{1}{r|}{$-5.34$} & \multicolumn{1}{r|}{$-12.16$} & \multicolumn{1}{r|}{$-1.22$} & \multicolumn{1}{r|}{$-1.27$} & \multicolumn{1}{r|}{$-1.25$} & \multicolumn{1}{r|}{$-0.91$} & \multicolumn{1}{r|}{$-1.25$} & \multicolumn{1}{r}{$-1.25$} \\ \hline
\multicolumn{1}{c|}{\multirow{3}{*}{$\ \,36$}} & \multicolumn{1}{c|}{$3\lambda $} & \multicolumn{1}{r|}{$-21.40$} & \multicolumn{1}{r|}{$-17.70$} & \multicolumn{1}{r|}{$-6.44$} & \multicolumn{1}{r|}{$-19.89$} & \multicolumn{1}{r|}{$-27.84$} & \multicolumn{1}{r|}{$-14.24$} & \multicolumn{1}{r|}{$-4.98$} & \multicolumn{1}{r|}{$-4.64$} & \multicolumn{1}{r}{$-3.00$} \\ \cline{2-11}
\multicolumn{1}{c|}{} & \multicolumn{1}{c|}{$6\lambda $} & \multicolumn{1}{r|}{$-16.43$} & \multicolumn{1}{r|}{$-51.35$} & \multicolumn{1}{r|}{$-77.43$} & \multicolumn{1}{r|}{$-3.98$} & \multicolumn{1}{r|}{$-7.35$} & \multicolumn{1}{r|}{$-3.94$} & \multicolumn{1}{r|}{$-2.12$} & \multicolumn{1}{r|}{$-2.42$} & \multicolumn{1}{r}{$-1.29$} \\ \cline{2-11}
\multicolumn{1}{c|}{} & \multicolumn{1}{c|}{$9\lambda $} & \multicolumn{1}{r|}{$-12.16$} & \multicolumn{1}{r|}{$-21.44$} & \multicolumn{1}{r|}{$-45.99$} & \multicolumn{1}{r|}{$-2.11$} & \multicolumn{1}{r|}{$-2.44$} & \multicolumn{1}{r|}{$-3.88$} & \multicolumn{1}{r|}{$-1.40$} & \multicolumn{1}{r|}{$-1.36$} & \multicolumn{1}{r}{$-1.25$} \\ \hline
\multicolumn{1}{c|}{\multirow{3}{*}{$\ \,81$}} & \multicolumn{1}{c|}{$3\lambda $} & \multicolumn{1}{r|}{$-32.90$} & \multicolumn{1}{r|}{$-19.59$} & \multicolumn{1}{r|}{$-5.42$} & \multicolumn{1}{r|}{$-20.93$} & \multicolumn{1}{r|}{$-15.51$} & \multicolumn{1}{r|}{$-32.51$} & \multicolumn{1}{r|}{$-11.39$} & \multicolumn{1}{r|}{$-8.93$} &\multicolumn{1}{r}{$-4.22$} \\ \cline{2-11}
\multicolumn{1}{c|}{} & \multicolumn{1}{c|}{$6\lambda $} & \multicolumn{1}{r|}{$-34.65$} & \multicolumn{1}{r|}{\color{blue}{$\bf{-186.34}$}} & \multicolumn{1}{r|}{\color{blue}{$\bf{-174.09}$}} & \multicolumn{1}{r|}{$-11.17$} & \multicolumn{1}{r|}{$-21.12$} & \multicolumn{1}{r|}{$-11.03$} & \multicolumn{1}{r|}{$-4.02$} & \multicolumn{1}{r|}{$-6.64$} & \multicolumn{1}{r}{$-5.23$} \\ \cline{2-11}
\multicolumn{1}{c|}{} & \multicolumn{1}{c|}{$9\lambda $} & \multicolumn{1}{r|}{$-20.34$} & \multicolumn{1}{r|}{\color{blue}{$\bf{-183.78}$}} & \multicolumn{1}{r|}{$-149.94$} & \multicolumn{1}{r|}{$-4.40$} & \multicolumn{1}{r|}{$-7.17$} & \multicolumn{1}{r|}{$-11.21$} & \multicolumn{1}{r|}{$-1.80$} & \multicolumn{1}{r|}{$-3.32$} & \multicolumn{1}{r}{$-2.81$} \\ \hline\hline
\multicolumn{11}{c}{The second-round experiment with moderate granularity} \\ \hline
\multicolumn{1}{c|}{\multirow{2}{*}{$M$}} &\multicolumn{1}{c|}{\multirow{2}{*}{$T_{\textrm{SIM}}$}} & \multicolumn{3}{c|}{$s_{\textrm{x}}=s_{\textrm{y}}=2\lambda $} & \multicolumn{3}{c|}{$s_{\textrm{x}}=s_{\textrm{y}}=2\lambda /3$} & \multicolumn{3}{c}{$s_{\textrm{x}}=s_{\textrm{y}}=2\lambda /5$} \\ \cline{3-11} 
\multicolumn{1}{c|}{} &\multicolumn{1}{c|}{} & \multicolumn{1}{c|}{$L=4$} & \multicolumn{1}{c|}{$L=6$} & \multicolumn{1}{c|}{$L=8$} & \multicolumn{1}{c|}{$L=4$} & \multicolumn{1}{c|}{$L=6$} & \multicolumn{1}{c|}{$L=8$} & \multicolumn{1}{c|}{$L=4$} & \multicolumn{1}{c|}{$L=6$} & $L=8$ \\ \hline
\multicolumn{1}{c|}{\multirow{3}{*}{$\ \,49$}} & \multicolumn{1}{c|}{$4\lambda $} & \multicolumn{1}{r|}{$-1.58$} & \multicolumn{1}{r|}{$-0.56$} & \multicolumn{1}{r|}{$-0.38$} & \multicolumn{1}{r|}{$-38.69$} & \multicolumn{1}{r|}{$-27.79$} & \multicolumn{1}{r|}{$-19.27$} & \multicolumn{1}{r|}{$-22.74$} & \multicolumn{1}{r|}{$-67.44$} & \multicolumn{1}{r}{$-19.13$} \\ \cline{2-11}
\multicolumn{1}{c|}{} & \multicolumn{1}{c|}{$6\lambda $} & \multicolumn{1}{r|}{$-8.24$} & \multicolumn{1}{r|}{$-2.11$} & \multicolumn{1}{r|}{$-0.83$} & \multicolumn{1}{r|}{$-21.11$} & \multicolumn{1}{r|}{$-64.99$} & \multicolumn{1}{r|}{$-41.89$} & \multicolumn{1}{r|}{$-13.64$} & \multicolumn{1}{r|}{$-13.34$} & \multicolumn{1}{r}{$-41.31$} \\ \cline{2-11}
\multicolumn{1}{c|}{} & \multicolumn{1}{c|}{$8\lambda $} & \multicolumn{1}{r|}{$-23.36$} & \multicolumn{1}{r|}{$-13.06$} & \multicolumn{1}{r|}{$-2.18$} & \multicolumn{1}{r|}{$-21.03$} & \multicolumn{1}{r|}{$-39.62$} & \multicolumn{1}{r|}{$-50.57$} & \multicolumn{1}{r|}{$-6.39$} & \multicolumn{1}{r|}{$-10.69$} & \multicolumn{1}{r}{$-15.80$} \\ \hline
\multicolumn{1}{c|}{\multirow{3}{*}{$\ \,81$}} & \multicolumn{1}{c|}{$4\lambda $} & \multicolumn{1}{r|}{$-1.66$} & \multicolumn{1}{r|}{$-0.52$} & \multicolumn{1}{r|}{$-0.38$} & \multicolumn{1}{r|}{$-39.88$} & \multicolumn{1}{r|}{$-27.21$} & \multicolumn{1}{r|}{$-28.59$} & \multicolumn{1}{r|}{$-39.46$} & \multicolumn{1}{r|}{$-49.97$} & \multicolumn{1}{r}{$-143.56$} \\ \cline{2-11}
\multicolumn{1}{c|}{} & \multicolumn{1}{c|}{$6\lambda $} & \multicolumn{1}{r|}{$-9.47$} & \multicolumn{1}{r|}{$-2.48$} & \multicolumn{1}{r|}{$-0.96$} & \multicolumn{1}{r|}{$-40.76$} & \multicolumn{1}{r|}{\color{blue}{$\bf{-186.34}$}} & \multicolumn{1}{r|}{$-55.88$} & \multicolumn{1}{r|}{$-23.20$} & \multicolumn{1}{r|}{\color{blue}{$\bf{-176.10}$}} & \multicolumn{1}{r}{$-20.38$} \\ \cline{2-11}
\multicolumn{1}{c|}{} & \multicolumn{1}{c|}{$8\lambda $} & \multicolumn{1}{r|}{$-21.78$} & \multicolumn{1}{r|}{$-5.61$} & \multicolumn{1}{r|}{$-3.37$} & \multicolumn{1}{r|}{$-31.07$} & \multicolumn{1}{r|}{$-71.25$} & \multicolumn{1}{r|}{\color{blue}{$\bf{-182.64}$}} & \multicolumn{1}{r|}{$-11.90$} & \multicolumn{1}{r|}{$-33.63$} & \multicolumn{1}{r}{$-9.54$} \\ \hline
\multicolumn{1}{c|}{\multirow{3}{*}{$121$}} & \multicolumn{1}{c|}{$4\lambda $} & \multicolumn{1}{r|}{$-1.28$} & \multicolumn{1}{r|}{$-0.54$} & \multicolumn{1}{r|}{$-0.36$} & \multicolumn{1}{r|}{$-32.92$} & \multicolumn{1}{r|}{$-74.72$} & \multicolumn{1}{r|}{$-16.65$} & \multicolumn{1}{r|}{\color{blue}{$\bf{-183.27}$}} & \multicolumn{1}{r|}{$-115.42$} & \multicolumn{1}{r}{\color{blue}{$\bf{-182.88}$}} \\ \cline{2-11}
\multicolumn{1}{c|}{} & \multicolumn{1}{c|}{$6\lambda $} & \multicolumn{1}{r|}{$-10.29$} & \multicolumn{1}{r|}{$-2.46$} & \multicolumn{1}{r|}{$-1.40$} & \multicolumn{1}{r|}{$-62.48$} & \multicolumn{1}{r|}{\color{blue}{$\bf{-179.98}$}} & \multicolumn{1}{r|}{\color{blue}{$\bf{-179.26}$}} & \multicolumn{1}{r|}{$-45.93$} & \multicolumn{1}{r|}{$-96.94$} & \multicolumn{1}{r}{\color{blue}{$\bf{-199.67}$}} \\ \cline{2-11}
\multicolumn{1}{c|}{} & \multicolumn{1}{c|}{$8\lambda $} & \multicolumn{1}{r|}{$-24.44$} & \multicolumn{1}{r|}{$-8.73$} & \multicolumn{1}{r|}{$-3.35$} & \multicolumn{1}{r|}{$-61.87$} & \multicolumn{1}{r|}{\color{blue}{$\bf{-199.91}$}} & \multicolumn{1}{r|}{\color{blue}{$\bf{-192.93}$}} & \multicolumn{1}{r|}{$-28.65$} & \multicolumn{1}{r|}{\color{blue}{$\bf{-194.52}$}} & \multicolumn{1}{r}{$-35.18$} \\ \hline\hline 
\multicolumn{11}{c}{The third-round experiment with fine granularity} \\ \hline
\multicolumn{1}{c|}{\multirow{2}{*}{$M$}} &\multicolumn{1}{c|}{\multirow{2}{*}{$T_{\textrm{SIM}}$}} & \multicolumn{3}{c|}{$s_{\textrm{x}}=s_{\textrm{y}}=2\lambda /2$} & \multicolumn{3}{c|}{$s_{\textrm{x}}=s_{\textrm{y}}=2\lambda /3$} & \multicolumn{3}{c}{$s_{\textrm{x}}=s_{\textrm{y}}=2\lambda /4$} \\ \cline{3-11} 
\multicolumn{1}{c|}{} &\multicolumn{1}{c|}{} & \multicolumn{1}{c|}{$L=5$} & \multicolumn{1}{c|}{$L=6$} & \multicolumn{1}{c|}{$L=7$} & \multicolumn{1}{c|}{$L=5$} & \multicolumn{1}{c|}{$L=6$} & \multicolumn{1}{c|}{$L=7$} & \multicolumn{1}{c|}{$L=5$} & \multicolumn{1}{c|}{$L=6$} & $L=7$ \\ \hline
\multicolumn{1}{c|}{\multirow{3}{*}{$100$}} & \multicolumn{1}{c|}{$7\lambda $} & \multicolumn{1}{r|}{$-34.33$} & \multicolumn{1}{r|}{$-31.57$} & \multicolumn{1}{r|}{$-40.62$} & \multicolumn{1}{r|}{$-52.49$} & \multicolumn{1}{r|}{\color{blue}{$\bf{-183.68}$}} & \multicolumn{1}{r|}{\color{blue}{$\bf{-185.46}$}} & \multicolumn{1}{r|}{$-78.29$} & \multicolumn{1}{r|}{\color{blue}{$\bf{-174.16}$}} & \multicolumn{1}{r}{$-65.66$} \\ \cline{2-11}
\multicolumn{1}{c|}{} & \multicolumn{1}{c|}{$8\lambda $} & \multicolumn{1}{r|}{\color{blue}{$\bf{-181.78}$}} & \multicolumn{1}{r|}{$-141.26$} & \multicolumn{1}{r|}{$-65.18$} & \multicolumn{1}{r|}{$-47.77$} & \multicolumn{1}{r|}{\color{blue}{$\bf{-190.77}$}} & \multicolumn{1}{r|}{$-100.66$} & \multicolumn{1}{r|}{\color{blue}{$\bf{-194.17}$}} & \multicolumn{1}{r|}{$-114.11$} & \color{blue}{$\bf{-182.10}$} \\ \cline{2-11}
\multicolumn{1}{c|}{} & \multicolumn{1}{c|}{$9\lambda $} & \multicolumn{1}{r|}{$-75.05$} & \multicolumn{1}{r|}{\color{blue}{$\bf{-186.58}$}} & \multicolumn{1}{r|}{$-28.49$} & \multicolumn{1}{r|}{$-52.09$} & \multicolumn{1}{r|}{$-59.48$} & \multicolumn{1}{r|}{\color{blue}{$\bf{-188.36}$}} & \multicolumn{1}{r|}{$-40.13$} & \multicolumn{1}{r|}{$-68.04$} & \color{blue}{$\bf{-192.96}$} \\ \hline
\multicolumn{1}{c|}{\multirow{3}{*}{$121$}} & \multicolumn{1}{c|}{$7\lambda $} & \multicolumn{1}{r|}{$-43.44$} & \multicolumn{1}{r|}{$-36.41$} & \multicolumn{1}{r|}{$-17.11$} & \multicolumn{1}{r|}{$-66.08$} & \multicolumn{1}{r|}{\color{blue}{$\bf{-188.02}$}} & \multicolumn{1}{r|}{\color{blue}{$\bf{-181.77}$}} & \multicolumn{1}{r|}{\color{blue}{$\bf{-194.05}$}} & \multicolumn{1}{r|}{\color{blue}{$\bf{-188.23}$}} & \color{blue}{$\bf{-187.96}$} \\ \cline{2-11}
\multicolumn{1}{c|}{} & \multicolumn{1}{c|}{$8\lambda $} & \multicolumn{1}{r|}{$-72.48$} & \multicolumn{1}{r|}{$-82.82$} & \multicolumn{1}{r|}{\color{blue}{$\bf{-180.05}$}} & \multicolumn{1}{r|}{$-78.40$} & \multicolumn{1}{r|}{\color{blue}{$\bf{-199.91}$}} & \multicolumn{1}{r|}{\color{blue}{$\bf{-194.52}$}} & \multicolumn{1}{r|}{$-93.94$} & \multicolumn{1}{r|}{\color{blue}{$\bf{-192.73}$}} & \color{blue}{$\bf{-177.78}$} \\ \cline{2-11}
\multicolumn{1}{c|}{} & \multicolumn{1}{c|}{$9\lambda $} & \multicolumn{1}{r|}{$-165.68$} & \multicolumn{1}{r|}{$-103.64$} & \multicolumn{1}{r|}{\color{blue}{$\bf{-185.65}$}} & \multicolumn{1}{r|}{$-39.28$} & \multicolumn{1}{r|}{$-78.45$} & \multicolumn{1}{r|}{\color{blue}{$\bf{-183.62}$}} & \multicolumn{1}{r|}{$-117.50$} & \multicolumn{1}{r|}{\color{blue}{$\bf{-183.12}$}} & \color{blue}{$\bf{-208.78}$} \\ \hline
\multicolumn{1}{c|}{\multirow{3}{*}{$144$}} & \multicolumn{1}{c|}{$7\lambda $} & \multicolumn{1}{r|}{$-35.95$} & \multicolumn{1}{r|}{$-163.67$} & \multicolumn{1}{r|}{$-34.67$} & \multicolumn{1}{r|}{\color{blue}{$\bf{-195.45}$}} & \multicolumn{1}{r|}{\color{blue}{$\bf{-191.91}$}} & \multicolumn{1}{r|}{\color{blue}{$\bf{-192.13}$}} & \multicolumn{1}{r|}{\color{blue}{$\bf{-186.43}$}} & \multicolumn{1}{r|}{\color{blue}{$\bf{-188.46}$}} & \color{blue}{$\bf{-179.55}$} \\ \cline{2-11}
\multicolumn{1}{c|}{} & \multicolumn{1}{c|}{$8\lambda $} & \multicolumn{1}{r|}{$-84.74$} & \multicolumn{1}{r|}{\color{blue}{$\bf{-181.21}$}} & \multicolumn{1}{r|}{$-91.63$} & \multicolumn{1}{r|}{$-72.60$} & \multicolumn{1}{r|}{\color{blue}{$\bf{-183.46}$}} & \multicolumn{1}{r|}{\color{blue}{$\bf{-201.35}$}} & \multicolumn{1}{r|}{$-52.73$} & \multicolumn{1}{r|}{\color{blue}{$\bf{-183.36}$}} & \color{blue}{$\bf{-178.34}$} \\ \cline{2-11}
\multicolumn{1}{c|}{} & \multicolumn{1}{c|}{$9\lambda $} & \multicolumn{1}{r|}{\color{blue}{$\bf{-183.27}$}} & \multicolumn{1}{r|}{$-105.71$} & \multicolumn{1}{r|}{\color{blue}{$\bf{-186.88}$}} & \multicolumn{1}{r|}{$-111.27$} & \multicolumn{1}{r|}{\color{blue}{$\bf{-174.52}$}} & \multicolumn{1}{r|}{\color{blue}{$\bf{-199.73}$}} & \multicolumn{1}{r|}{$-44.56$} & \multicolumn{1}{r|}{\color{blue}{$\bf{-180.33}$}} & \color{blue}{$\bf{-178.95}$} \\ \hline
\end{tabular}\vspace{-0.6cm}
\end{table*}

In this section, we report on our numerical simulations to evaluate the performance of the SIM-based DOA estimator. The system layout of estimating the DOA is shown in Fig. 1. The system operates at $60$ GHz. Unless otherwise specified, we consider a square UPA having $N_{\textrm{x}} \times N_{\textrm{y}}$ elements. The array spacing is set to $d_{\textrm{x}}=d_{\textrm{y}}=\lambda /2$.
\subsection{Ablation Study}
First, we conduct extensive simulations to examine the optimal SIM designed for implementing 2D DFT having grid points of $\left ( 2,2 \right )$ and $\left ( 4,4 \right )$, respectively. Specifically, a SIM has four key parameters: \emph{i)} The thickness $T_{\textrm{SIM}}$ of the SIM; \emph{ii)} The number $L$ of metasurface layers; \emph{iii)} The number $M$ of meta-atoms per layer; and \emph{iv)} The spacing between elements in the $x$ and $y$ directions, namely $s_{\textrm{x}}$ and $s_{\textrm{y}}$. For brevity, we characterize a SIM using a four-tuple $\left ( T_{\textrm{SIM}}, L, M, s_{\textrm{x}} \right )$, assuming $s_{\textrm{x}} = s_{\textrm{y}}$. Since testing over all four parameters across many possible values would be extremely time-consuming, we design a three-round ablation study having successively improved granularity. Taking the number of layers $L$ as an example, in the first round we consider $L=3,6,9$ along with a step size of $3$. After identifying the best SIM configuration, we narrow the search range around the optimal value and reduce the step size to $2$. Similarly, the step size is further reduced to $1$ in the third round. The other parameters are listed in the following tables. For brevity, we assume $u_{\textrm{x}} = u_{\textrm{y}} = \lambda/2$.

Table \ref{tab1} shows the results of a SIM fitting the 2D DFT matrix having $\left ( 2, 2 \right )$ grids. All results are obtained by averaging $100$ independent experiments. The optimal solution found in the first round of experiments is used as the center point for the second-round experiment associated with moderate granularity. The same process is repeated for the third-round experiment. After the three rounds, it is observed that the optimal SIM designed for fitting the 2D DFT matrix having $\left ( 2, 2 \right )$ grids is $\left ( 9\lambda,\,7,\,121,\,2\lambda/4 \right )$, achieving a normalized MSE of $-208.78$ dB. Furthermore, we also mark those SIM setups that achieve satisfactory results (defined as $\mathcal{L} \leq -170$ dB) in bold blue font. As the experiments progress, more SIM setups are examined to achieve NMSE values lower than the target. This verifies that the optimal SIM setup is not exclusively determined and a practical system needs to adaptively design the SIM. 

Moreover, Table \ref{tab2} shows the results for the 2D DFT having $\left ( 4,4 \right )$ grid points. Those SIM setups that achieve an NMSE less than $-15$ dB are marked in bold blue font. After three rounds of experiments, the optimal SIM design for the 2D DFT having $\left ( 4,4 \right )$ grid points is found to be around $\left ( 12\lambda,\,13,\,225,\,4\lambda/9 \right )$. It is important to note that due to the challenge of fitting a larger 2D DFT matrix, the fitting NMSE in Table \ref{tab2} is higher than that in Table \ref{tab1}. Nevertheless, later we will demonstrate that this has a negligible effect on the DOA estimation performance.

Based on the results in Tables \ref{tab1} and \ref{tab2}, there exist fundamental tradeoffs between the four parameters of the SIM. More specifically, we summarize our pivotal findings as follows:
\begin{itemize}
 \item For a fixed number of layers $L$, the inter-layer propagation matrix $\boldsymbol{W}_{l}$ may become singular, as the thickness of the SIM $T_{\textrm{SIM}}$ increases without limit, while a very thin SIM causes $\boldsymbol{W}_{l}$ to become nearly diagonal, both lacking the diversity for accurately fitting the 2D DFT matrix.
 \item For a fixed thickness $T_{\textrm{SIM}}$, increasing the number of layers $L$ results in a denser metasurface arrangement, rendering $\boldsymbol{W}_{l}$ nearly diagonal. Too few metasurface layers may lack the adequate DoF to leverage all wave propagation components in the SIM.
 \item An excessive number of meta-atoms would result in unnecessary links within the SIM and even introduce adverse wave propagation components to fit the 2D DFT matrix. However, too few meta-atoms may violate \emph{Theorem \ref{theorem1}} and restrict the SIM to fit $\boldsymbol{F}$ accurately.
 \item For an excessive element spacing $s_{\textrm{x}}$, the inter-layer propagation matrix $\boldsymbol{W}_{l}$ tends towards diagonal, while a low element spacing may result in identical values across all entries, resulting in a rank-one matrix unable to fit the 2D DFT matrix of rank $N$.
\end{itemize}
Therefore, we should carefully design the SIM for practical applications. The rigorous evaluation of the fitting capability of SIM may involve complex matrix decomposition theory, warranting future efforts.
\begin{table*}[!t]
\centering
\scriptsize
\caption{The normalized loss function value (in dB) of utilizing SIM to fit 2D DFT matrix with $\left ( 4,4 \right )$ grid points.}
\label{tab2}
\begin{tabular}{ccccccccccc}
\hline
\multicolumn{11}{c}{The first-round experiment with coarse granularity} \\ \hline
\multicolumn{1}{c|}{\multirow{2}{*}{$M$}} &\multicolumn{1}{c|}{\multirow{2}{*}{$T_{\textrm{SIM}}$}} & \multicolumn{3}{c|}{$s_{\textrm{x}}=s_{\textrm{y}}=4\lambda /6$} & \multicolumn{3}{c|}{$s_{\textrm{x}}=s_{\textrm{y}}=4\lambda /9$} & \multicolumn{3}{c}{$s_{\textrm{x}}=s_{\textrm{y}}=4\lambda /12$} \\ \cline{3-11}
\multicolumn{1}{c|}{} & \multicolumn{1}{c|}{} & \multicolumn{1}{c|}{$L=6$} & \multicolumn{1}{c|}{$L=9$} & \multicolumn{1}{c|}{$L=12$} & \multicolumn{1}{c|}{$L=6$} & \multicolumn{1}{c|}{$L=9$} & \multicolumn{1}{c|}{$L=12$} & \multicolumn{1}{c|}{$L=6$} & \multicolumn{1}{c|}{$L=9$} & $L=12$ \\ \hline
\multicolumn{1}{c|}{\multirow{3}{*}{$\ \, 36$}} & \multicolumn{1}{c|}{$\ \, 6\lambda $} & \multicolumn{1}{r|}{$-3.90$} & \multicolumn{1}{r|}{$-4.07$} & \multicolumn{1}{r|}{$-3.81$} & \multicolumn{1}{r|}{$-3.13$} & \multicolumn{1}{r|}{$-4.36$} & \multicolumn{1}{r|}{$-4.76$} & \multicolumn{1}{r|}{$-1.58$} & \multicolumn{1}{r|}{$-1.75$} & \multicolumn{1}{r}{$-1.32$} \\ \cline{2-11}
\multicolumn{1}{c|}{} & \multicolumn{1}{c|}{$\ \, 9\lambda $} & \multicolumn{1}{r|}{$-4.27$} & \multicolumn{1}{r|}{$-5.95$} & \multicolumn{1}{r|}{$-6.22$} & \multicolumn{1}{r|}{$-2.04$} & \multicolumn{1}{r|}{$-3.00$} & \multicolumn{1}{r|}{$-3.42$} & \multicolumn{1}{r|}{$-1.24$} & \multicolumn{1}{r|}{$-1.45$} & \multicolumn{1}{r}{$-1.11$} \\ \cline{2-11}
\multicolumn{1}{c|}{} & \multicolumn{1}{c|}{$12\lambda $} & \multicolumn{1}{r|}{$-3.43$} & \multicolumn{1}{r|}{$-6.14$} & \multicolumn{1}{r|}{$-6.71$} & \multicolumn{1}{r|}{$-1.91$} & \multicolumn{1}{r|}{$-2.19$} & \multicolumn{1}{r|}{$-2.13$} & \multicolumn{1}{r|}{$-0.88$} & \multicolumn{1}{r|}{$-1.04$} & \multicolumn{1}{r}{$-0.59$} \\ \hline
\multicolumn{1}{c|}{\multirow{3}{*}{$\ \,81$}} & \multicolumn{1}{c|}{$\ \, 6\lambda $} & \multicolumn{1}{r|}{$-7.65$} & \multicolumn{1}{r|}{$-7.97$} & \multicolumn{1}{r|}{$-4.55$} & \multicolumn{1}{r|}{$-8.26$} & \multicolumn{1}{r|}{$-11.80$} & \multicolumn{1}{r|}{$-11.24$} & \multicolumn{1}{r|}{$-5.00$} & \multicolumn{1}{r|}{$-5.62$} & \multicolumn{1}{r}{$-5.15$} \\ \cline{2-11}
\multicolumn{1}{c|}{} & \multicolumn{1}{c|}{$\ \, 9\lambda $} & \multicolumn{1}{r|}{$-9.21$} & \multicolumn{1}{r|}{$-10.82$} & \multicolumn{1}{r|}{$-10.06$} & \multicolumn{1}{r|}{$-6.64$} & \multicolumn{1}{r|}{$-10.02$} & \multicolumn{1}{r|}{$-10.83$} & \multicolumn{1}{r|}{$-3.71$} & \multicolumn{1}{r|}{$-3.94$} & \multicolumn{1}{r}{$-2.74$} \\ \cline{2-11}
\multicolumn{1}{c|}{} & \multicolumn{1}{c|}{$12\lambda $} & \multicolumn{1}{r|}{$-8.71$} & \multicolumn{1}{r|}{$-11.69$} & \multicolumn{1}{r|}{$-11.13$} & \multicolumn{1}{r|}{$-5.12$} & \multicolumn{1}{r|}{$-8.15$} & \multicolumn{1}{r|}{$-9.44$} & \multicolumn{1}{r|}{$-2.46$} & \multicolumn{1}{r|}{$-2.83$} & \multicolumn{1}{r}{$-2.72$} \\ \hline
\multicolumn{1}{c|}{\multirow{3}{*}{$144$}} & \multicolumn{1}{c|}{$\ \, 6\lambda $} & \multicolumn{1}{r|}{$-11.03$} & \multicolumn{1}{r|}{$-10.41$} & \multicolumn{1}{r|}{$-5.83$} & \multicolumn{1}{r|}{$-13.60$} & \multicolumn{1}{r|}{$-15.94$} & \multicolumn{1}{r|}{$-15.41$} & \multicolumn{1}{r|}{$-10.73$} & \multicolumn{1}{r|}{$-11.31$} &\multicolumn{1}{r}{$-6.99$} \\ \cline{2-11}
\multicolumn{1}{c|}{} & \multicolumn{1}{c|}{$\ \, 9\lambda $} & \multicolumn{1}{r|}{$-12.38$} & \multicolumn{1}{r|}{$-15.41$} & \multicolumn{1}{r|}{$-13.55$} & \multicolumn{1}{r|}{$-12.65$} & \multicolumn{1}{r|}{$-17.24$} & \multicolumn{1}{r|}{$-18.01$} & \multicolumn{1}{r|}{$-8.32$} & \multicolumn{1}{r|}{$-9.08$} & \multicolumn{1}{r}{$-9.12$} \\ \cline{2-11}
\multicolumn{1}{c|}{} & \multicolumn{1}{c|}{$12\lambda $} & \multicolumn{1}{r|}{$-12.20$} & \multicolumn{1}{r|}{$-15.09$} & \multicolumn{1}{r|}{$-15.87$} & \multicolumn{1}{r|}{$-9.66$} & \multicolumn{1}{r|}{$-14.86$} & \multicolumn{1}{r|}{$-15.91$} & \multicolumn{1}{r|}{$-6.12$} & \multicolumn{1}{r|}{$-7.93$} & \multicolumn{1}{r}{$-8.38$} \\ \hline\hline
\multicolumn{11}{c}{The second-round experiment with moderate granularity} \\ \hline
\multicolumn{1}{c|}{\multirow{2}{*}{$M$}} &\multicolumn{1}{c|}{\multirow{2}{*}{$T_{\textrm{SIM}}$}} & \multicolumn{3}{c|}{$s_{\textrm{x}}=s_{\textrm{y}}=4\lambda /7$} & \multicolumn{3}{c|}{$s_{\textrm{x}}=s_{\textrm{y}}=4\lambda /9$} & \multicolumn{3}{c}{$s_{\textrm{x}}=s_{\textrm{y}}=4\lambda /11$} \\ \cline{3-11}
\multicolumn{1}{c|}{} &\multicolumn{1}{c|}{} & \multicolumn{1}{c|}{$L=10$} & \multicolumn{1}{c|}{$L=12$} & \multicolumn{1}{c|}{$L=14$} & \multicolumn{1}{c|}{$L=10$} & \multicolumn{1}{c|}{$L=12$} & \multicolumn{1}{c|}{$L=14$} & \multicolumn{1}{c|}{$L=10$} & \multicolumn{1}{c|}{$L=12$} & $L=14$ \\ \hline
\multicolumn{1}{c|}{\multirow{3}{*}{$100$}} & \multicolumn{1}{c|}{$\ \,7\lambda $} & \multicolumn{1}{r|}{$-15.16$} & \multicolumn{1}{r|}{$-14.05$} & \multicolumn{1}{r|}{$-14.25$} & \multicolumn{1}{r|}{$-12.63$} & \multicolumn{1}{r|}{$-13.66$} & \multicolumn{1}{r|}{$-13.82$} & \multicolumn{1}{r|}{$-9.42$} & \multicolumn{1}{r|}{$-8.45$} & \multicolumn{1}{r}{$-8.49$} \\ \cline{2-11}
\multicolumn{1}{c|}{} & \multicolumn{1}{c|}{$\ \,9\lambda $} & \multicolumn{1}{r|}{$-14.44$} & \multicolumn{1}{r|}{$-15.22$} & \multicolumn{1}{r|}{$-14.62$} & \multicolumn{1}{r|}{$-12.40$} & \multicolumn{1}{r|}{$-12.02$} & \multicolumn{1}{r|}{$-14.92$} & \multicolumn{1}{r|}{$-7.83$} & \multicolumn{1}{r|}{$-6.44$} & \multicolumn{1}{r}{$-3.23$} \\ \cline{2-11}
\multicolumn{1}{c|}{} & \multicolumn{1}{c|}{$11\lambda $} & \multicolumn{1}{r|}{$-14.35$} & \multicolumn{1}{r|}{$-15.21$} & \multicolumn{1}{r|}{$-15.28$} & \multicolumn{1}{r|}{$-11.96$} & \multicolumn{1}{r|}{$-11.69$} & \multicolumn{1}{r|}{$-12.18$} & \multicolumn{1}{r|}{$-6.71$} & \multicolumn{1}{r|}{$-4.90$} & \multicolumn{1}{r}{$-4.00$} \\ \hline
\multicolumn{1}{c|}{\multirow{3}{*}{$144$}} & \multicolumn{1}{c|}{$\ \,7\lambda $} & \multicolumn{1}{r|}{$-15.31$} & \multicolumn{1}{r|}{$-16.06$} & \multicolumn{1}{r|}{$-15.45$} & \multicolumn{1}{r|}{$-17.70$} & \multicolumn{1}{r|}{$-19.31$} & \multicolumn{1}{r|}{$-17.58$} & \multicolumn{1}{r|}{$-13.06$} & \multicolumn{1}{r|}{$-12.42$} & \multicolumn{1}{r}{$-8.87$} \\ \cline{2-11}
\multicolumn{1}{c|}{} & \multicolumn{1}{c|}{$\ \,9\lambda $} & \multicolumn{1}{r|}{$-14.38$} & \multicolumn{1}{r|}{$-17.73$} & \multicolumn{1}{r|}{$-17.49$} & \multicolumn{1}{r|}{$-17.88$} & \multicolumn{1}{r|}{$-18.01$} & \multicolumn{1}{r|}{$-17.93$} & \multicolumn{1}{r|}{$-14.08$} & \multicolumn{1}{r|}{$-12.55$} & \multicolumn{1}{r}{$-10.60$} \\ \cline{2-11}
\multicolumn{1}{c|}{} & \multicolumn{1}{c|}{$11\lambda $} & \multicolumn{1}{r|}{$-15.51$} & \multicolumn{1}{r|}{$-17.38$} & \multicolumn{1}{r|}{$-18.73$} & \multicolumn{1}{r|}{$-16.42$} & \multicolumn{1}{r|}{$-17.26$} & \multicolumn{1}{r|}{\color{blue}{$\bf{-19.75}$}} & \multicolumn{1}{r|}{$-11.81$} & \multicolumn{1}{r|}{$-10.89$} & \multicolumn{1}{r}{$-6.55$} \\ \hline
\multicolumn{1}{c|}{\multirow{3}{*}{$196$}} & \multicolumn{1}{c|}{$\ \,7\lambda $} & \multicolumn{1}{r|}{$-15.73$} & \multicolumn{1}{r|}{$-15.81$} & \multicolumn{1}{r|}{$-16.27$} & \multicolumn{1}{r|}{\color{blue}{$\bf{-20.95}$}} & \multicolumn{1}{r|}{$-17.34$} & \multicolumn{1}{r|}{$-19.12$} & \multicolumn{1}{r|}{$-16.46$} & \multicolumn{1}{r|}{$-17.67$} & \multicolumn{1}{r}{$-16.87$} \\ \cline{2-11}
\multicolumn{1}{c|}{} & \multicolumn{1}{c|}{$\ \,9\lambda $} & \multicolumn{1}{r|}{$-16.43$} & \multicolumn{1}{r|}{$-17.37$} & \multicolumn{1}{r|}{\color{blue}{$\bf{-20.88}$}} & \multicolumn{1}{r|}{$-19.07$} & \multicolumn{1}{r|}{\color{blue}{$\bf{-21.38}$}} & \multicolumn{1}{r|}{$-19.48$} & \multicolumn{1}{r|}{$-15.01$} & \multicolumn{1}{r|}{$-17.81$} & \multicolumn{1}{r}{$-14.50$} \\ \cline{2-11}
\multicolumn{1}{c|}{} & \multicolumn{1}{c|}{$11\lambda $} & \multicolumn{1}{r|}{$-18.23$} & \multicolumn{1}{r|}{$-19.42$} & \multicolumn{1}{r|}{$-19.43$} & \multicolumn{1}{r|}{\color{blue}{$\bf{-19.94}$}} & \multicolumn{1}{r|}{\color{blue}{$\bf{-22.20}$}} & \multicolumn{1}{r|}{\color{blue}{$\bf{-21.04}$}} & \multicolumn{1}{r|}{$-16.34$} & \multicolumn{1}{r|}{$-17.60$} & \multicolumn{1}{r}{$-16.35$} \\ \hline\hline
\multicolumn{11}{c}{The third-round experiment with fine granularity} \\ \hline
\multicolumn{1}{c|}{\multirow{2}{*}{$M$}} &\multicolumn{1}{c|}{\multirow{2}{*}{$T_{\textrm{SIM}}$}} & \multicolumn{3}{c|}{$s_{\textrm{x}}=s_{\textrm{y}}=4\lambda /8$} & \multicolumn{3}{c|}{$s_{\textrm{x}}=s_{\textrm{y}}=4\lambda /9$} & \multicolumn{3}{c}{$s_{\textrm{x}}=s_{\textrm{y}}=4\lambda /10$} \\ \cline{3-11} 
\multicolumn{1}{c|}{} &\multicolumn{1}{c|}{} & \multicolumn{1}{c|}{$L=11$} & \multicolumn{1}{c|}{$L=12$} & \multicolumn{1}{c|}{$L=13$} & \multicolumn{1}{c|}{$L=11$} & \multicolumn{1}{c|}{$L=12$} & \multicolumn{1}{c|}{$L=13$} & \multicolumn{1}{c|}{$L=11$} & \multicolumn{1}{c|}{$L=12$} & $L=13$ \\ \hline
\multicolumn{1}{c|}{\multirow{3}{*}{$169$}} & \multicolumn{1}{c|}{$10\lambda $} & \multicolumn{1}{r|}{$-19.48$} & \multicolumn{1}{r|}{$-17.72$} & \multicolumn{1}{r|}{$-17.97$} & \multicolumn{1}{r|}{$-16.58$} & \multicolumn{1}{r|}{$-19.43$} & \multicolumn{1}{r|}{\color{blue}{$\bf{-22.67}$}} & \multicolumn{1}{r|}{$-15.92$} & \multicolumn{1}{r|}{$-16.34$} & \multicolumn{1}{r}{$-17.18$} \\ \cline{2-11}
\multicolumn{1}{c|}{} & \multicolumn{1}{c|}{$11\lambda $} & \multicolumn{1}{r|}{$-17.69$} & \multicolumn{1}{r|}{\color{blue}{$\bf{-19.86}$}} & \multicolumn{1}{r|}{\color{blue}{$\bf{-20.27}$}} & \multicolumn{1}{r|}{$-16.04$} & \multicolumn{1}{r|}{\color{blue}{$\bf{-19.74}$}} & \multicolumn{1}{r|}{$-17.21$} & \multicolumn{1}{r|}{$-17.31$} & \multicolumn{1}{r|}{$-18.26$} & \multicolumn{1}{r}{$-14.77$} \\ \cline{2-11}
\multicolumn{1}{c|}{} & \multicolumn{1}{c|}{$12\lambda $} & \multicolumn{1}{r|}{$-18.47$} & \multicolumn{1}{r|}{$-17.58$} & \multicolumn{1}{r|}{\color{blue}{$\bf{-21.00}$}} & \multicolumn{1}{r|}{$-18.26$} & \multicolumn{1}{r|}{\color{blue}{$\bf{-19.80}$}} & \multicolumn{1}{r|}{$-18.34$} & \multicolumn{1}{r|}{$-13.74$} & \multicolumn{1}{r|}{$-15.71$} & \multicolumn{1}{r}{$-14.44$} \\ \hline
\multicolumn{1}{c|}{\multirow{3}{*}{$196$}} & \multicolumn{1}{c|}{$10\lambda $} & \multicolumn{1}{r|}{$-18.17$} & \multicolumn{1}{r|}{\color{blue}{$\bf{-21.06}$}} & \multicolumn{1}{r|}{\color{blue}{$\bf{-22.02}$}} & \multicolumn{1}{r|}{\color{blue}{$\bf{-20.05}$}} & \multicolumn{1}{r|}{\color{blue}{$\bf{-24.15}$}} & \multicolumn{1}{r|}{\color{blue}{$\bf{-20.46}$}} & \multicolumn{1}{r|}{\color{blue}{$\bf{-20.03}$}} & \multicolumn{1}{r|}{$-18.22$} & \multicolumn{1}{r}{$-19.07$} \\ \cline{2-11}
\multicolumn{1}{c|}{} & \multicolumn{1}{c|}{$11\lambda $} & \multicolumn{1}{r|}{$-18.87$} & \multicolumn{1}{r|}{$-19.10$} & \multicolumn{1}{r|}{$-18.82$} & \multicolumn{1}{r|}{$-17.20$} & \multicolumn{1}{r|}{\color{blue}{$\bf{-22.20}$}} & \multicolumn{1}{r|}{\color{blue}{$\bf{-20.37}$}} & \multicolumn{1}{r|}{$-17.49$} & \multicolumn{1}{r|}{\color{blue}{$\bf{-19.92}$}} & \multicolumn{1}{r}{$-16.18$} \\ \cline{2-11}
\multicolumn{1}{c|}{} & \multicolumn{1}{c|}{$12\lambda $} & \multicolumn{1}{r|}{\color{blue}{$\bf{-20.52}$}} & \multicolumn{1}{r|}{$-18.69$} & \multicolumn{1}{r|}{\color{blue}{$\bf{-20.56}$}} & \multicolumn{1}{r|}{\color{blue}{$\bf{-21.07}$}} & \multicolumn{1}{r|}{\color{blue}{$\bf{-20.95}$}} & \multicolumn{1}{r|}{$-19.14$} & \multicolumn{1}{r|}{$-19.42$} & \multicolumn{1}{r|}{$-18.69$} & \multicolumn{1}{r}{$-17.27$} \\ \hline
\multicolumn{1}{c|}{\multirow{3}{*}{$225$}} & \multicolumn{1}{c|}{$10\lambda $} & \multicolumn{1}{r|}{\color{blue}{$\bf{-19.65}$}} & \multicolumn{1}{r|}{\color{blue}{$\bf{-20.32}$}} & \multicolumn{1}{r|}{\color{blue}{$\bf{-20.24}$}} & \multicolumn{1}{r|}{\color{blue}{$\bf{-21.69}$}} & \multicolumn{1}{r|}{\color{blue}{$\bf{-20.79}$}} & \multicolumn{1}{r|}{\color{blue}{$\bf{-22.17}$}} & \multicolumn{1}{r|}{$-17.59$} & \multicolumn{1}{r|}{\color{blue}{$\bf{-20.13}$}} & \multicolumn{1}{r}{$-18.84$} \\ \cline{2-11}
\multicolumn{1}{c|}{} & \multicolumn{1}{c|}{$11\lambda $} & \multicolumn{1}{r|}{\color{blue}{$\bf{-20.23}$}} & \multicolumn{1}{r|}{\color{blue}{$\bf{-19.69}$}} & \multicolumn{1}{r|}{\color{blue}{$\bf{-22.92}$}} & \multicolumn{1}{r|}{\color{blue}{$\bf{-22.16}$}} & \multicolumn{1}{r|}{\color{blue}{$\bf{-19.80}$}} & \multicolumn{1}{r|}{\color{blue}{$\bf{-22.52}$}} & \multicolumn{1}{r|}{$-19.28$} & \multicolumn{1}{r|}{$-19.62$} & \multicolumn{1}{r}{\color{blue}{$\bf{-20.61}$}} \\ \cline{2-11}
\multicolumn{1}{c|}{} & \multicolumn{1}{c|}{$12\lambda $} & \multicolumn{1}{r|}{\color{blue}{$\bf{-19.94}$}} & \multicolumn{1}{r|}{\color{blue}{$\bf{-21.87}$}} & \multicolumn{1}{r|}{\color{blue}{$\bf{-20.15}$}} & \multicolumn{1}{r|}{\color{blue}{$\bf{-20.58}$}} & \multicolumn{1}{r|}{$-18.60$} & \multicolumn{1}{r|}{\color{blue}{$\bf{-24.17}$}} & \multicolumn{1}{r|}{\color{blue}{$\bf{-21.41}$}} & \multicolumn{1}{c|}{\color{blue}{$\bf{-20.27}$}} & \multicolumn{1}{r}{\color{blue}{$\bf{-19.90}$}} \\ \hline
\end{tabular}\vspace{-0.6cm}
\end{table*}
\subsection{Convergence Behavior of the Proposed Gradient Descent Algorithm}
\begin{figure}[t!]
	\centering
	\subfloat[\label{fig_5a}The normalized loss function value $\mathcal{L}$ versus the number of iterations;]{
		\includegraphics[width=6cm]{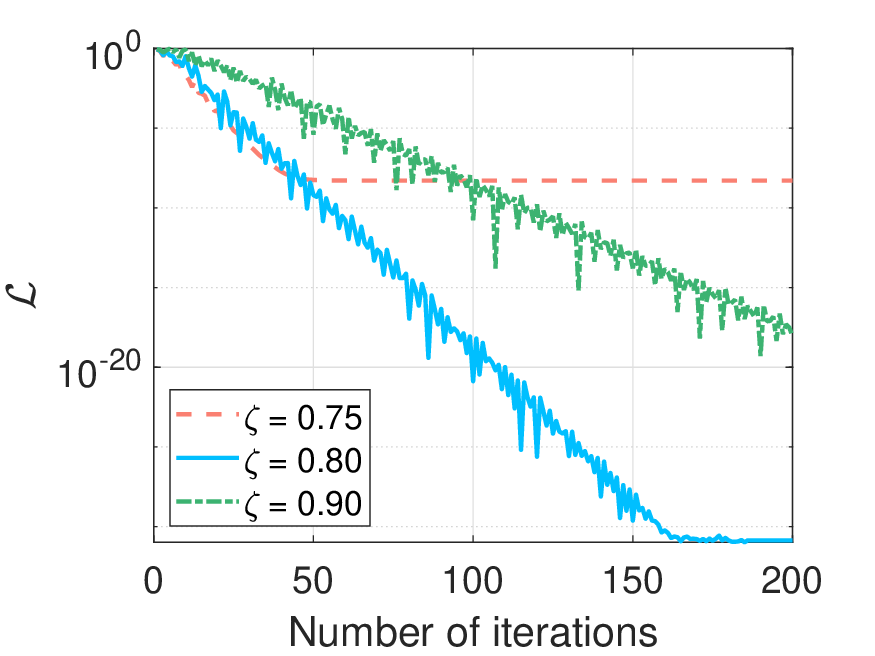}}\quad
	\subfloat[\label{fig_5b}The normalized loss function value $\mathcal{L}$ versus the decay parameter $\zeta$.]{
		\includegraphics[width=6cm]{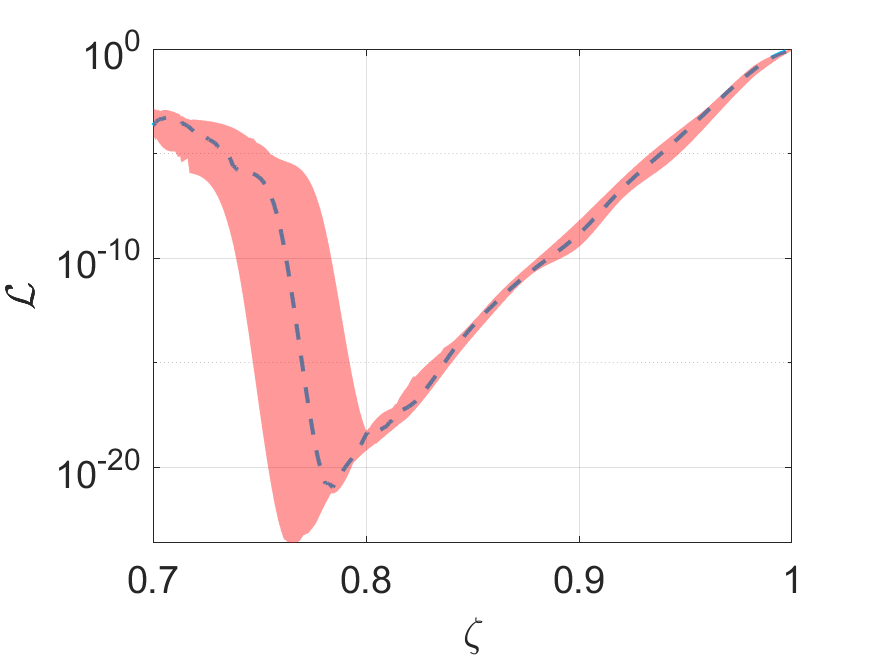}}
 	\caption{The convergence behavior of the proposed gradient descent algorithm for optimizing a SIM to fit a 2D DFT matrix with $\left ( 2,2 \right )$ grid points.}\vspace{-0.4cm}
	\label{fig_5} 
\end{figure}

Fig. \ref{fig_5} evaluates the convergence behavior of the proposed gradient descent algorithm for optimizing a SIM to fit a 2D DFT matrix of $\left ( 2,2 \right )$ grid points. According to Table \ref{tab1}, the optimal SIM is fabricated by inserting $L = 7$ layers into a cube having a thickness of $T_{\text{SIM}} = 9\lambda$. Each square metasurface contains $M = 121$ meta-atoms associated with $s_{\text{x}} = s_{\text{y}} = \lambda/2$ element spacing. First, Fig. \ref{fig_5a} plots the normalized loss function value $\mathcal{L}$ versus the number of iterations for three cases with different decay parameter values, namely $\zeta = 0.75,\, 0.80,\, 0.90$. It demonstrates that as the iterations proceed, the proposed gradient descent method gradually converges for a moderate decay parameter value, such as $0.80$. For a value of $\zeta$ close to $1$, the algorithm may overshoot frequently and need a higher number of iterations to converge. While the gradient descent method may reduce the loss function rapidly, namely within less than $50$ iterations for $\zeta = 0.75$, it may get stuck at a locally optimal point. To further demonstrate the effect of decay parameters, Fig. \ref{fig_5b} examines the resultant loss function value of $\mathcal{L}$ after $100$ iterations versus the decay parameter $\zeta$. Specifically, $50$ independent experiments are conducted, and both the average value and potential range (shown as the red zone) are plotted. It can be seen that as $\zeta$ increases, the SIM's fitting performance first improves and then degrades, which is consistent with our analysis. A decay parameter of about $\zeta = 0.80$ achieves the best performance under the worst-case condition.

\begin{figure}[t!]
	\centering
	\subfloat[\label{fig_6a}The normalized loss function value $\mathcal{L}$ versus the number of iterations;]{
		\includegraphics[width=6cm]{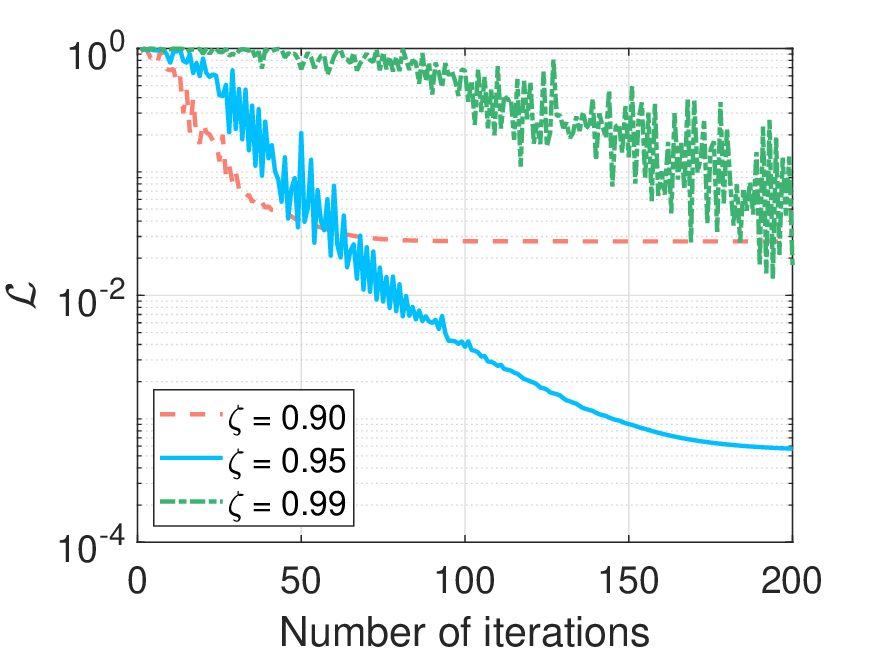}}\quad
	\subfloat[\label{fig_6b}The normalized loss function value $\mathcal{L}$ versus the decay parameter $\zeta$.]{
		\includegraphics[width=6cm]{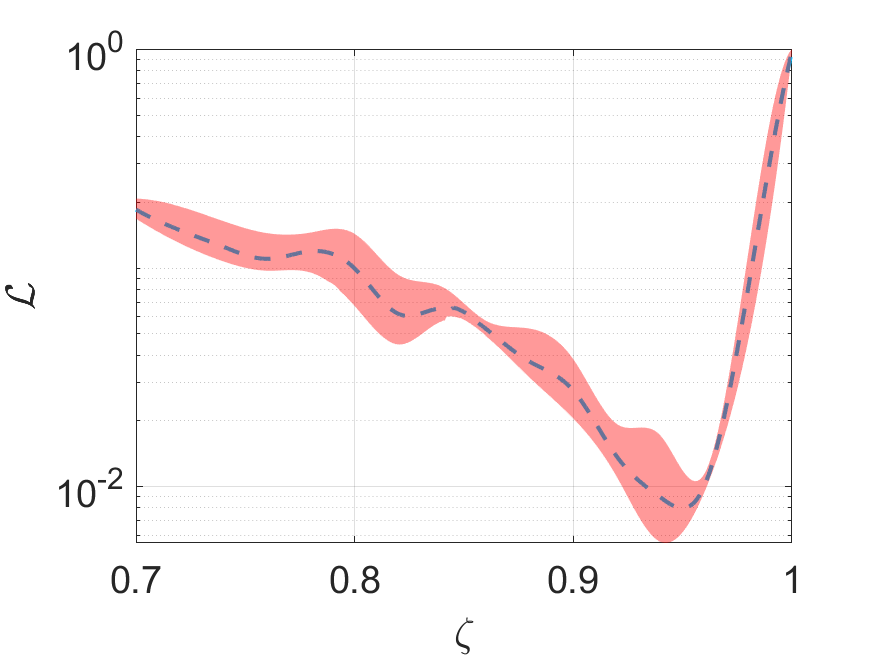}}
	\caption{The convergence behavior of the proposed gradient descent algorithm for optimizing a SIM to fit a 2D DFT matrix with $\left ( 4,4 \right )$ grid points.}\vspace{-0.6cm}
	\label{fig_6} 
\end{figure}

Furthermore, Fig. \ref{fig_6} verifies the convergence behavior of the gradient descent algorithm for leveraging a SIM to fit a 2D DFT matrix of $\left ( 4,4 \right )$ grids. The SIM hardware parameters are set to $\left ( 12\lambda ,\, 13,\, 225,\, 4\lambda/9 \right )$ based on the results in Table \ref{tab2}. Three different decay parameter values of $\zeta = 0.90,\, 0.95,\, 0.99$ are considered. As shown in Fig. \ref{fig_6a}, a moderate decay parameter value, such as $\zeta = 0.95$, is capable of accurately fitting the 2D DFT matrix at an NMSE of less than $10^{-3}$ after about $200$ iterations. Larger or smaller decay parameters would deteriorate the performance of the gradient descent method, and this trend is similar to Fig. \ref{fig_5a}. Again, due to the challenge of utilizing a SIM to fit the 2D DFT matrix with more grid points, the resultant fitting performance is poorer than that for the $\left ( 2, 2 \right )$-grid scenario, even though a larger SIM is used. Moreover, Fig. \ref{fig_6b} evaluates the effects of $\zeta$ in this scenario by running $50$ independent experiments. The optimal $\zeta$ that minimizes the normalized loss function value within $100$ iterations is about $\zeta = 0.95$. This validates the necessity to judiciously adjust the learning rate during iterations. In summary, in early iterations, a lower $\zeta$ would be preferred to allow for larger phase shift changes, and then it should be appropriately increased for maintaining a smooth learning rate update to fine-tune the SIM's response.

\subsection{2D DFT Capability of SIM}
\begin{figure}[!t]
	\centering
	\subfloat[\label{fig_7a}2D DFT in the wave domain;]{
		\includegraphics[width=6cm]{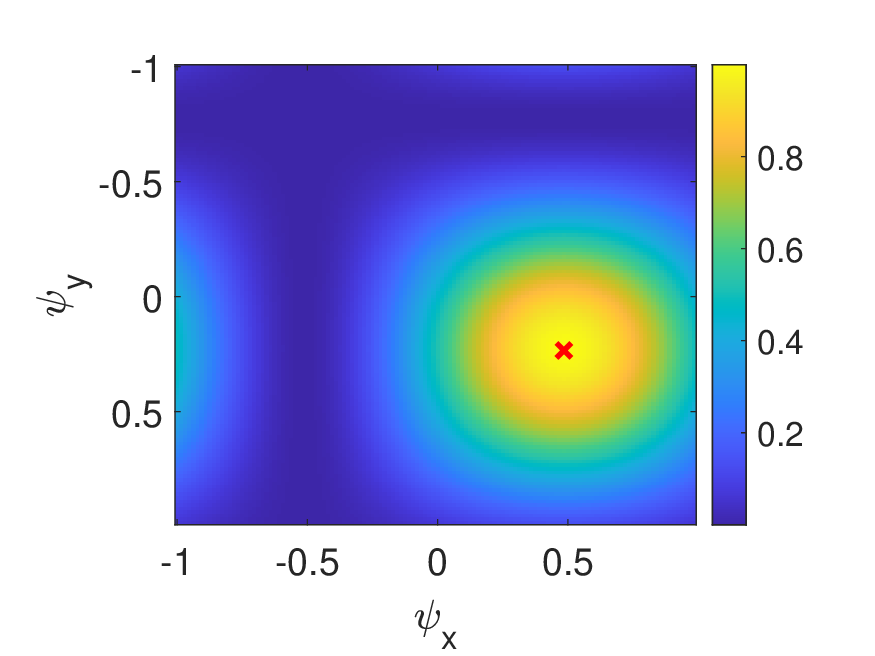}}
	\subfloat[\label{fig_7b}2D DFT in the digital domain.]{
		\includegraphics[width=6cm]{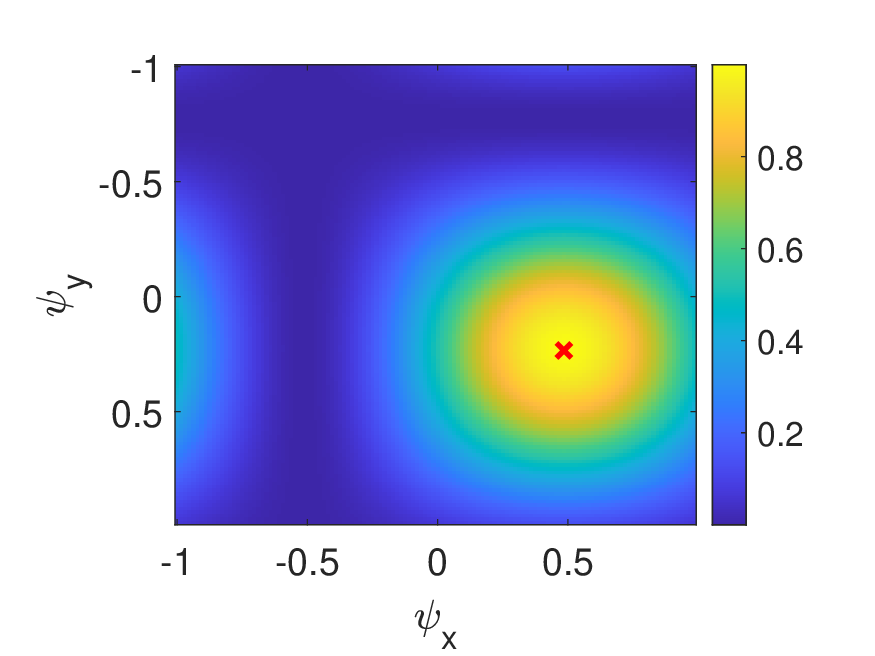}}
	\caption{Angular spectrum of the incoming signal using a $\left ( 2\times 2 \right )$ receiver array, where we have $\bar{\psi} _{\textrm{x}}=0.48$ and $\bar{\psi} _{\textrm{y}}=0.23$. The red cross represents the true DOA position on the spectrum.}\vspace{-0.4cm}
	\label{fig_7}
\end{figure}

To demonstrate the 2D DFT's capability in the SIM, we next examine the angular spectrum of the incoming signal by first considering the case of $N_{\textrm{x}} = N_{\textrm{y}} = 2$ and $T_{\textrm{x}} = T_{\textrm{y}} = 64$. Specifically, following the protocol outlined in Section \ref{sec4_1}, we collect the received signal passing through a well-trained SIM in each snapshot. The SIM has been pre-optimized using the gradient descent method along with $\zeta = 0.8$ over $200$ iterations. We then use the outputs from $T = 4096$ snapshots{\footnote{For the sake of illustration, here we use a large value of $T$, but later we will evaluate the DOA estimation performance of the SIM for a moderate number of snapshots.}} to generate the angular spectrum of the incoming signal, as shown in Fig. \ref{fig_7a}. The spectrum peak is normalized. Moreover, the electrical angles corresponding to the $x$ and $y$ directions are set to $\bar{\psi}_{\textrm{x}}=0.48$ and $\bar{\psi}_{\textrm{y}}=0.23$, respectively, as marked by a red cross in the figure. Note that the spectrum generated by the SIM succeeds in focusing the beam toward the true DOA position. This allows estimating the DOA parameters by simply searching for the antenna and snapshot having the strongest received magnitude. Furthermore, Fig. \ref{fig_7b} plots the spectrum using 2D DFT in the digital domain under the same protocol. Upon comparing Figs. \ref{fig_7a} and \ref{fig_7b}, we note that the SIM outputs almost the same signal spectrum as the 2D digital DFT. Nevertheless, the wave-based computing paradigm is totally different from the conventional digital approach. As the computations are carried out naturally as signals propagate through the SIM, the computing delay is significantly reduced and the hardware design can also be accordingly simplified.

Moreover, Figs. \ref{fig_8a} and \ref{fig_8b} show the angular spectra using the SIM and the 2D digital DFT, respectively, considering a $\left ( 4\times 4 \right )$ receiver array and $T_{\textrm{x}} = T_{\textrm{y}} = 32$. The angular spectrum of the incoming signal is generated by collecting the received signals passing through the SIM over $T = 1024$ snapshots. The SIM adopts the optimal hardware parameters listed in Table \ref{tab2} and it is optimized through $200$ iterations with a decay parameter of $\zeta = 0.95$. The electrical angles in the $x$ and $y$ directions are set to $\bar{\psi}_{\textrm{x}}=-0.58$ and $\bar{\psi}_{\textrm{y}}=-0.28$, respectively. Thanks to the increased array aperture, the angular spectra in Fig. \ref{fig_8} have reduced leakage compared to those in Fig. \ref{fig_7}. As such, one can obtain a more accurate DOA estimate under noiseless conditions. Again, we note that the SIM is capable of generating the same angular spectrum as the 2D digital DFT for the $\left ( 4\times 4 \right )$ array and perfectly concentrating the received signal onto the antenna and snapshot corresponding to the true DOA value.

\subsection{Validation of Theoretical Analysis}
Next, we verify the accuracy of our analytical results by examining the MSE of using a SIM for DOA estimation. For brevity, we estimate the electrical angles, which can be mapped to the true elevation and azimuth angles using a bijection. Firstly, Fig. \ref{fig_9a} plots the MSE versus the effective SNR, which is defined as $\varrho \beta ^{2}$. We also consider different numbers of snapshots: $T_{\textrm{x}} = T_{\textrm{y}} = 2$ and $T_{\textrm{x}} = T_{\textrm{y}} = 4$. For each setup, we perform $1,000$ independent experiments, where the DOA parameters of the single radiation source are uniformly distributed in the upper half-space of the SIM, as shown in Fig. \ref{fig_2}. The SIM hardware parameters are appropriately selected as in Table \ref{tab1}, while the phase shifts are optimized leveraging the gradient descent with $\zeta =0.8$. Moreover, the theoretical MSE values are obtained by averaging $1,000$ corresponding values calculated from \emph{Theorem 2}. As expected, the estimation accuracy improves for all scenarios, as the SNR increases. However, due to the discrete nature of the SIM, which can only return the on-grid estimate of DOA parameters, the proposed SIM-based estimator eventually reaches an error floor determined by the resolution in terms of angular bins. Nevertheless, the estimation performance can be further improved by collecting more snapshots. For example, increasing the number of snapshots from $T = 4$ to $T = 16$ provides about a $20$ dB gain in terms of the SNR. Additionally, the performance bound analytically derived from \emph{Theorem 2} serves as an upper bound in both cases. As the SNR increases, the gap between the simulation and analytical results gradually narrows, since the scaling in \eqref{eq48} becomes tight.

\begin{figure}[t!]
	\centering
	\subfloat[\label{fig_8a}2D DFT in the wave domain;]{
		\includegraphics[width=6cm]{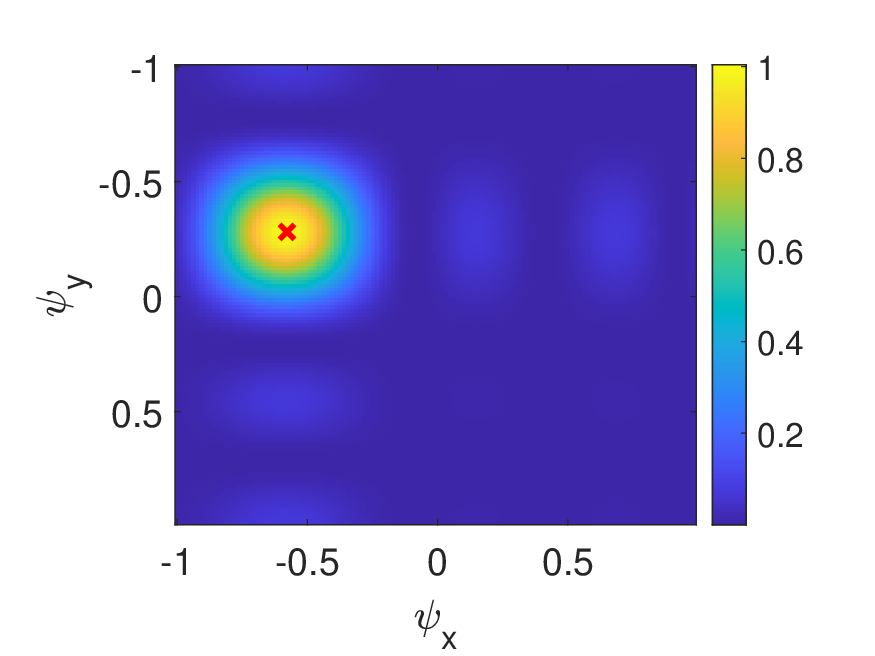}}
	\subfloat[\label{fig_8b}2D DFT in the digital domain.]{
		\includegraphics[width=6cm]{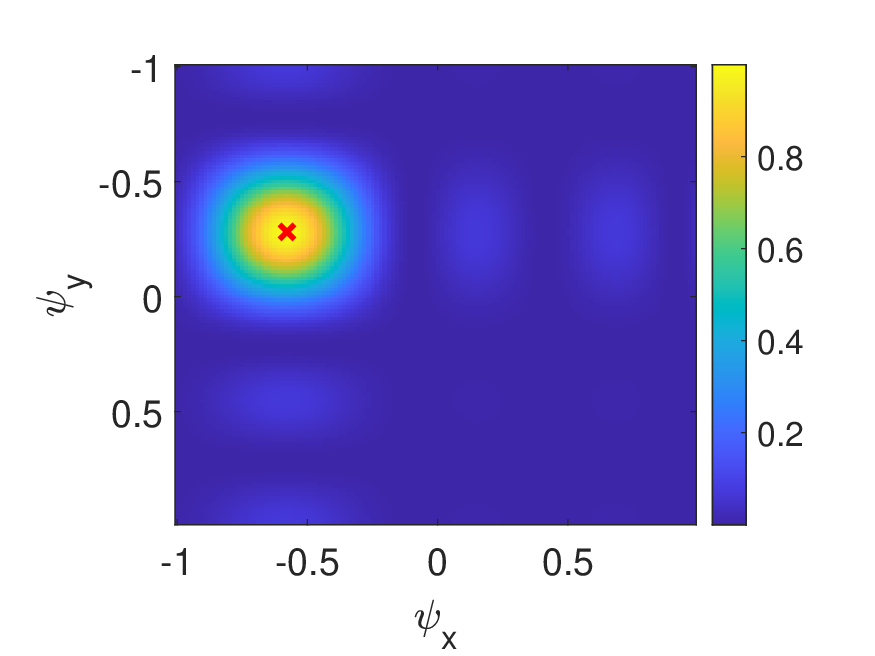}}
	\caption{Angular spectrum of the incoming signal using a $\left ( 4\times 4 \right )$ receiver array, where we have $\bar{\psi} _{\textrm{x}}=-0.58$ and $\bar{\psi}_{\textrm{y}}=-0.28$. The red cross represents the true DOA position on the spectrum.}\vspace{-0.6cm}
	\label{fig_8}
\end{figure}

Fig. \ref{fig_9b} evaluates the MSE of the SIM-based estimator versus the effective SNR, considering a $\left ( 4\times 4 \right )$ receiver array. The SIM's hardware parameters and phase shifts remain the same as in Fig. \ref{fig_8}. Additionally, two different numbers of snapshots are considered: $T_{\textrm{x}} = T_{\textrm{y}} = 4$ and $T_{\textrm{x}} = T_{\textrm{y}} = 8$. As seen in Fig. \ref{fig_9b}, both the theoretical and simulation results indicate that the MSE improves as the SNR and the number of snapshots increase. Specifically, when increasing the number of snapshots from $T = 16$ to $T=64$ at an effective SNR of $10$ dB, the MSE is reduced from $1.5\times 10^{-3}$ to $0.75\times10^{-3}$, resulting in a $3$ dB MSE improvement. Further increasing the effective SNR to $30$ dB provides an extra $3$ dB of performance gain. At high SNRs, the theoretical upper bound (i.e., green curves in Fig. \ref{fig_9}) becomes asymptotically tight. Moreover, compared to Fig. \ref{fig_9a} employing a $\left ( 2\times 2 \right )$ receiver array with $T_{\textrm{x}} = T_{\textrm{y}} = 4$, doubling the array aperture reduces the MSE from $0.6 \times 10^{-2}$ to $1.3 \times 10^{-3}$, providing about a $6$ dB gain in MSE.

\subsection{Performance Comparison with Conventional Beamforming Methods}
\begin{figure*}[t!]
	\centering
	\subfloat[\label{fig_9a}With a $\left ( 2\times 2 \right )$ receiver array;]{
		\includegraphics[width=7.5cm]{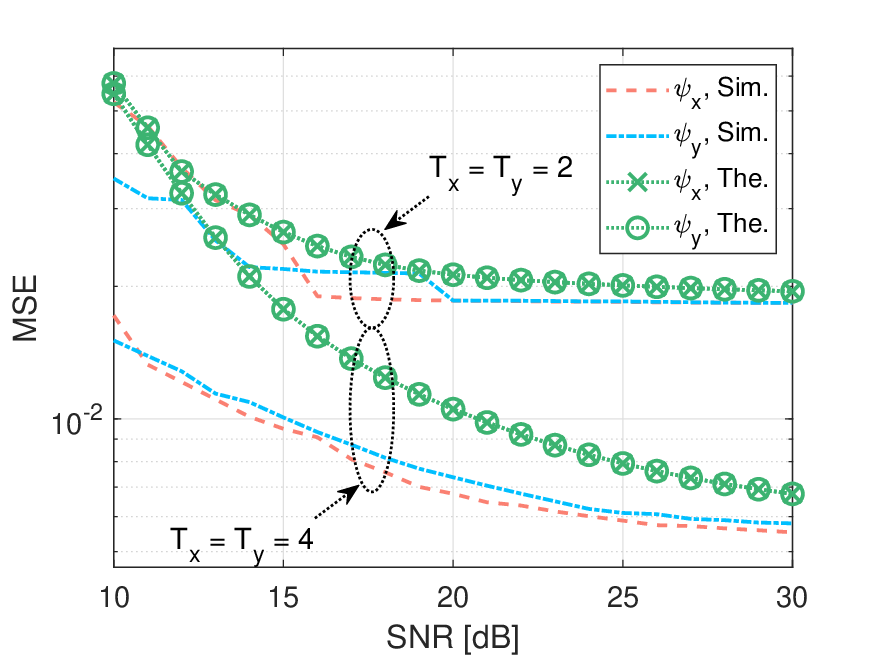}}
	\subfloat[\label{fig_9b}With a $\left ( 4\times 4 \right )$ receiver array.]{
		\includegraphics[width=7.5cm]{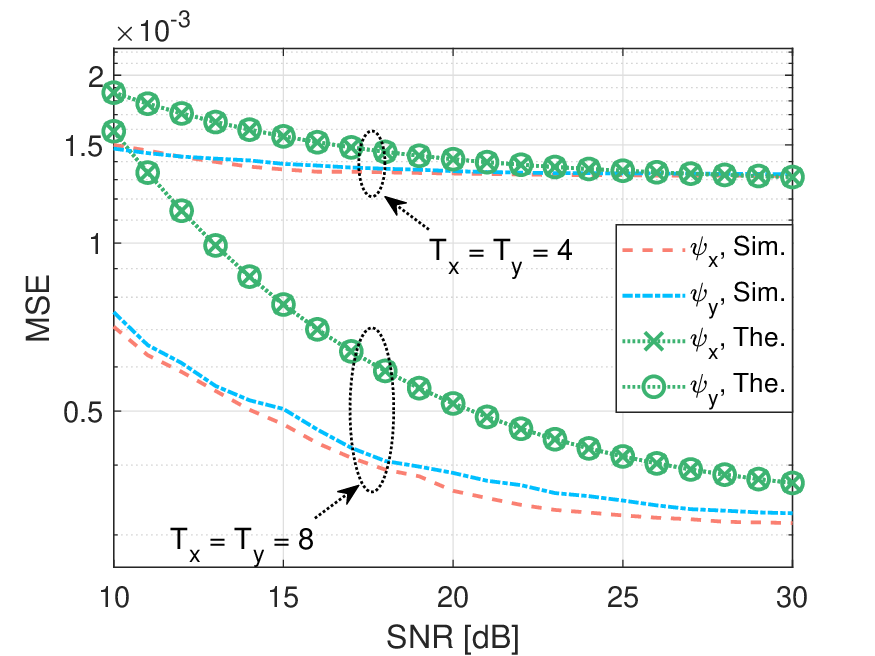}}
	\caption{The MSE of the SIM-based DOA estimator versus the effective SNR.}\vspace{-0.6cm}
	\label{fig_9} 
\end{figure*}

\begin{figure*}[t!]
	\centering
	\subfloat[\label{fig_10a}With a $\left ( 2\times 2 \right )$ receiver array;]{
		\includegraphics[width=7.5cm]{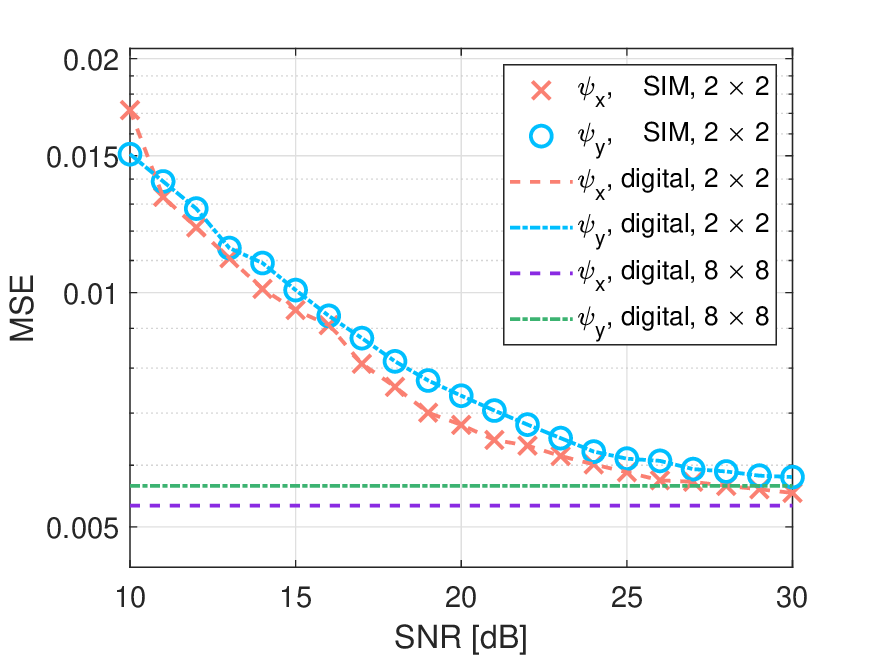}}
	\subfloat[\label{fig_10b}With a $\left ( 4\times 4 \right )$ receiver array.]{
		\includegraphics[width=7.5cm]{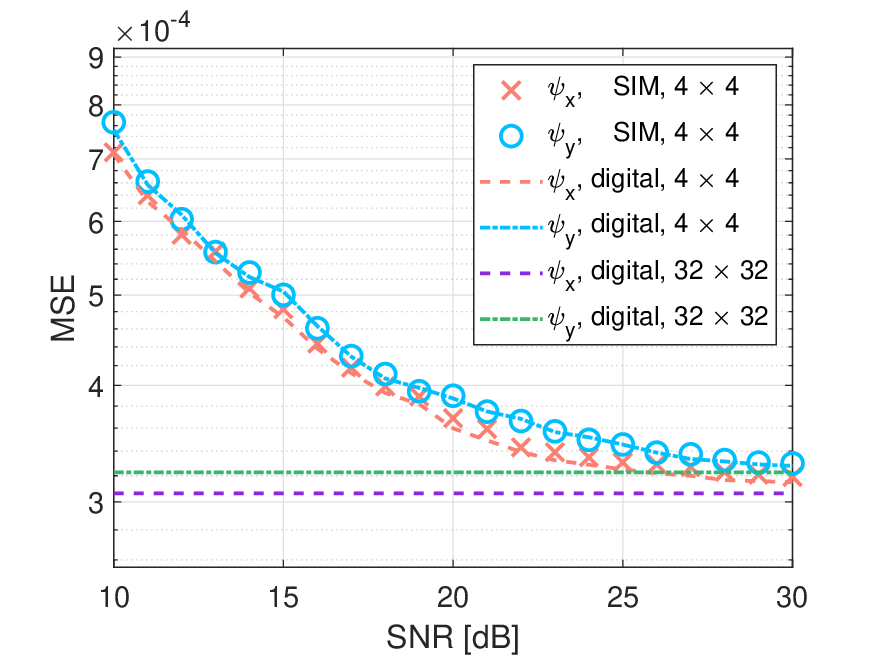}}
	\caption{Performance comparison of the SIM-based DOA estimator and the conventional approach via a 2D digital DFT.}\vspace{-0.6cm}
	\label{fig_10} 
\end{figure*}

Furthermore, we compare the performance of the proposed SIM-based DOA estimator to the conventional beamforming-based method, which matches the steering vector to an estimated array response. The simulation results shown in Fig. \ref{fig_10} consider the cases of $N_{\textrm{x}} = N_{\textrm{y}} = 2$ and $N_{\textrm{x}} = N_{\textrm{y}} = 4$, respectively, while keeping all other parameters the same as in Fig. \ref{fig_9}. Moreover, the corresponding numbers of snapshots are set to $T=16$ and $T=64$, respectively. Fig. \ref{fig_10} shows that the proposed SIM-based estimator performs similarly well to the digital beamforming-based method under all setups. However, the digital method requires phase-sensitive receivers and digital signal processing, while the SIM-based estimator, relies on energy detection and wave-based signal processing. Additionally, Fig. \ref{fig_10a} also plots the MSE of an $\left ( 8\times 8 \right )$ receiver array in a single snapshot, which provides a lower bound for the SIM-based scheme. Observe that the gap gradually narrows as the effective SNR increases. Similarly, the receiver array of $\left ( 32\times 32 \right )$ in a single snapshot characterizes the lower bound of the SIM-based estimator using a $\left ( 4\times 4 \right )$ receiver array and $T=64$ snapshots, as shown in Fig. \ref{fig_10b}. Although the digital method with a $\left ( 32\times 32 \right )$ array is capable of accurately estimating the DOA across all SNR regions, it requires a large array aperture, whereas the SIM-based system estimates the DOA with a smaller array and a moderate number of snapshots.
\subsection{Effects of Receiver Array Arrangement}
\begin{figure}[!t]
\centering
\includegraphics[width=12cm]{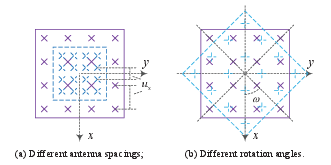}
\caption{A SIM-aided array system.}\vspace{-0.4cm}
\label{fig_11}
\end{figure}

\begin{figure}[t!]
	\centering
	\subfloat[\label{fig_12a}The normalized loss function $\mathcal{L}$ versus the antenna spacing $u_{\text{x}}$;]{
		\includegraphics[width=6cm]{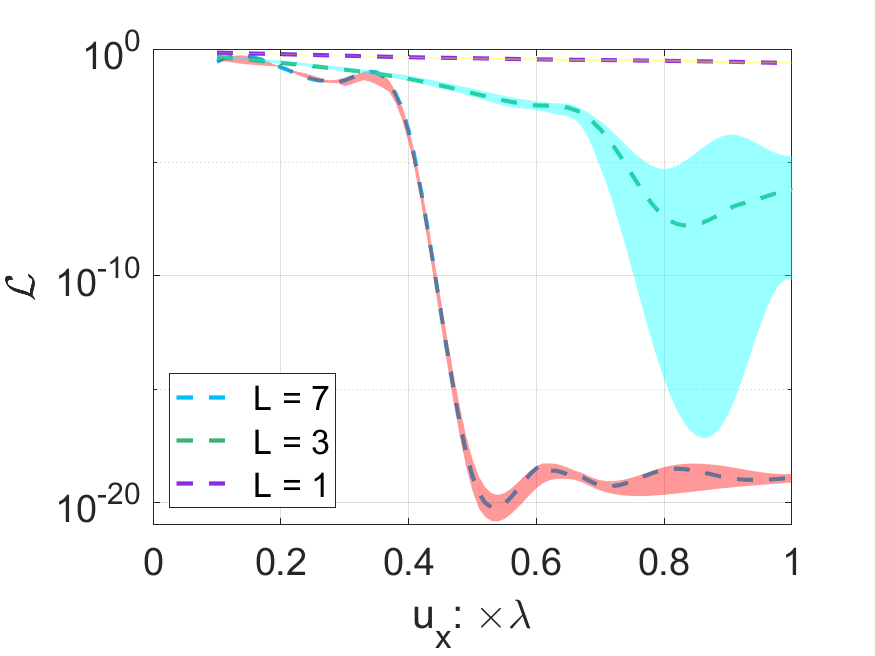}}\quad
	\subfloat[\label{fig_12b}The normalized loss function $\mathcal{L}$ versus the rotation angle $\omega $.]{
		\includegraphics[width=6cm]{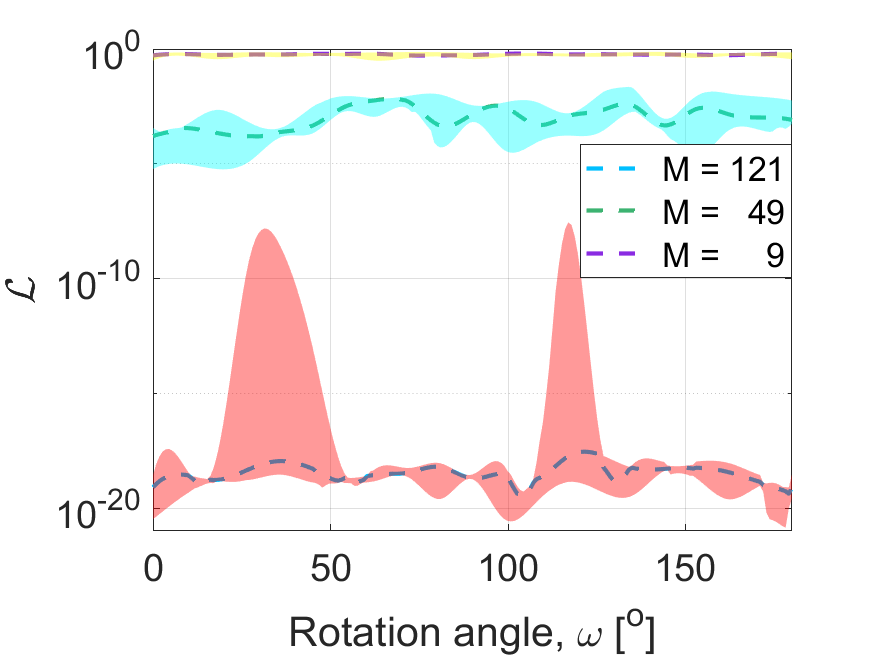}}
	\caption{The effects of the receiver array arrangement: a $2 \times 2$ array.}\vspace{-0.4cm}
	\label{fig_12} 
\end{figure}

Finally, we evaluate the effects of the receiver array arrangement on the SIM's capability of performing 2D DFT. As shown in Fig. \ref{fig_11}, we consider two main parameters: \emph{i)} the antenna spacing $u_\text{x} = u_\text{y}$; and \emph{ii)} the rotation angle $\omega$. Fig. \ref{fig_12a} shows the fitting NMSE versus the antenna spacing $u_\text{x}$, assuming $N_\text{x} = N_\text{y} = 2$ and $\omega = 0^{\circ}$. In each case, the SIM phase shifts are optimized by performing $100$ iterations with a decay parameter of $\zeta = 0.8$. The SIM's hardware parameters are the same as in Fig. \ref{fig_7}, while we also consider different numbers of layers: $L = 1,\,3,\,7$. The results are obtained by averaging $50$ independent experiments, with the ranges also plotted. Fig. \ref{fig_12a} demonstrates that a SIM having fewer layers cannot fit the 2D DFT matrix well. A SIM of seven metasurface layers is capable of accurately performing 2D DFT when $u_\text{x} \geq \lambda/2$. Furthermore, Fig. \ref{fig_12b} evaluates the effects of the rotation angle of the receiver array, considering $u_\text{x} = \lambda/2$ and different numbers of meta-atoms on each layer: $M=9,\,49,\,121$. Fig. \ref{fig_12b} shows that a SIM having a small number of meta-atoms lacks sufficient inference capability. For a SIM having $M=121$ meta-atoms per layer, the fitting performance becomes insensitive to the rotation angle of the receiver array. Nonetheless, for a rotation angle near $\omega =45^{\circ}$, the SIM using the proposed gradient descent may have poorer robustness. In summary, this verifies the effectiveness of our previous setups adopting $u_\text{x} = u_\text{y} = \lambda/2$ and $\omega =0^{\circ}$.

\begin{figure}[t!]
	\centering
	\subfloat[\label{fig_13a}The normalized loss function $\mathcal{L}$ versus the antenna spacing $u_{\text{x}}$;]{
		\includegraphics[width=6cm]{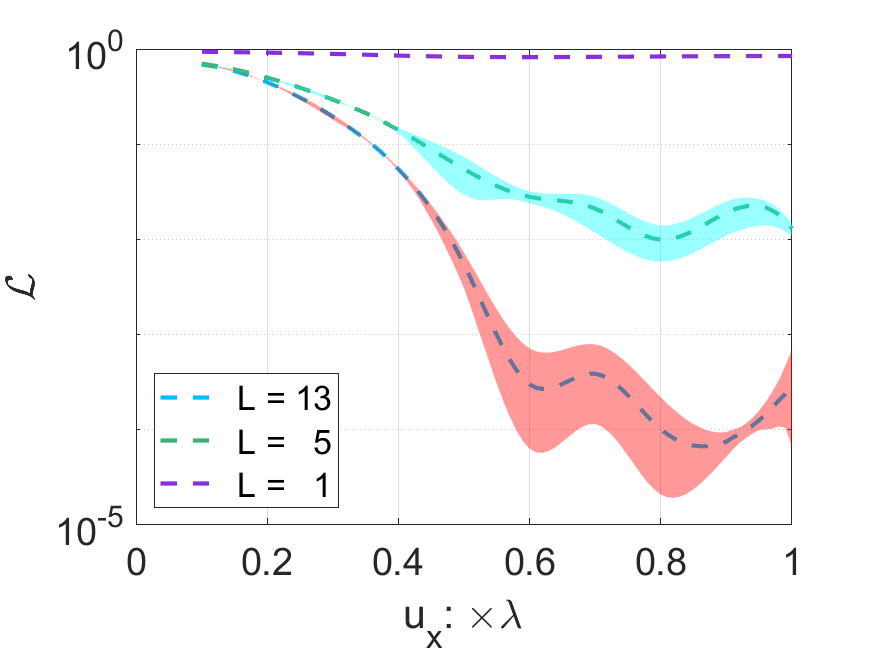}}\quad
	\subfloat[\label{fig_13b}The normalized loss function $\mathcal{L}$ versus the rotation angle $\omega $.]{
		\includegraphics[width=6cm]{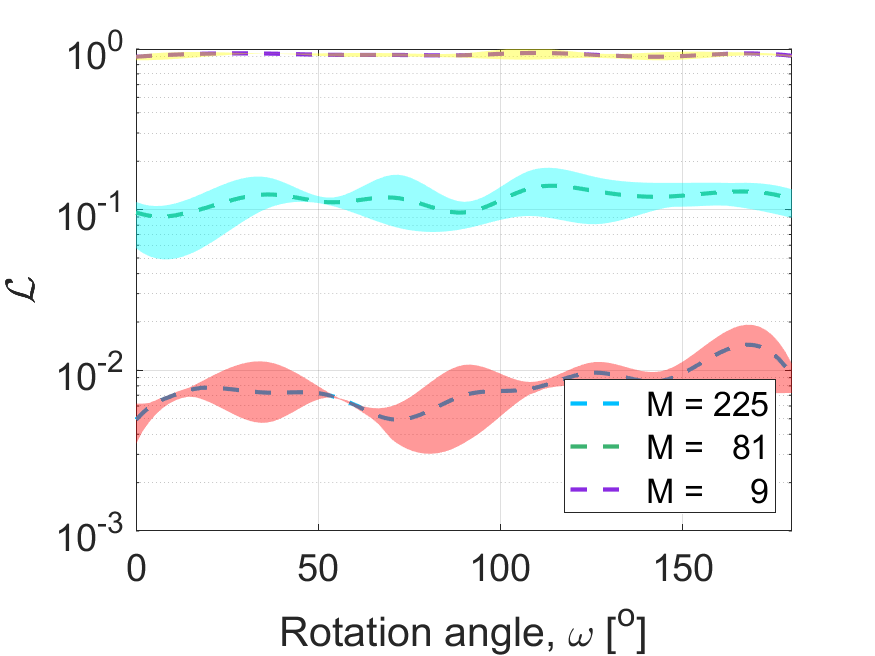}}
	\caption{The effects of the receiver array arrangement: a $4 \times 4$ array.}\vspace{-0.6cm}
	\label{fig_13} 
\end{figure}

Fig. \ref {fig_13a} evaluates the effects of the antenna spacing $u_\text{x}$ in the receiver array on the SIM's capability of performing 2D DFT with $\left ( 4,4 \right )$ grid points. The SIM has the same hardware parameters as in Fig. \ref{fig_8}, while the SIM phase shifts are optimized through $100$ iterations with a decay parameter of $\zeta = 0.95$. We consider different numbers of layers: $L = 1,\,5,\,13$. Similarly, a SIM with fewer layers cannot fit the 2D DFT matrix well, while a SIM with $L = 13$ metasurface layers is capable of accurately performing 2D DFT when $u_\text{x} \geq \lambda/2$. Moreover, Fig. \ref{fig_13b} evaluates the effects of the rotation angle $\omega$ of the receiver array, considering $u_\text{x} = \lambda/2$ and different numbers of meta-atoms on each layer: $M=9,\,81,\,225$. A small number of meta-atoms results in poor inference capability. For a SIM having an adequate number of meta-atoms (such as $M=225$) per layer, the fitting performance becomes less sensitive to $\omega$. Again, this verifies that the optimal receiver array should have an isomorphic arrangement with the input layer of the SIM.

\section{Conclusions}\label{sec7}
A novel SIM architecture has been proposed for estimating the 2D DOA parameters of a single radiation source. This new design allows for a significant improvement in computation speed and hardware complexity. By customizing a gradient descent method to guide the SIM to perform a 2D DFT, the spatial EM waves can be automatically transformed into their angular domain as they propagate through the SIM. Furthermore, we have developed a protocol to generate a set of angular spectra with orthogonal spatial frequency bins. As a result, the DOA can be estimated by searching for the antenna and snapshot having the strongest magnitude. Also, we theoretically evaluated and numerically verified the MSE of the proposed SIM-based DOA estimator. Extensive simulation results were provided to examine the optimal SIM hardware parameters and hyperparameters for the gradient descent method. In particular, we summarize our main findings as follows:
\begin{itemize}
 \item A SIM having $9\lambda$ thickness and $7$ metasurface layers, with $121$ $\lambda/2$-spaced meta-atoms per layer, is capable of minimizing the loss function for fitting the 2D DFT matrix of $\left ( 2,2 \right )$ grid points. For the $\left ( 4,4 \right )$-grid scenario, the optimal SIM design has a thickness of $12\lambda$ and $13$ layers, each with $225$ meta-atoms spaced $4\lambda/9$ apart.
 \item The optimal decay parameter for the gradient descent method when fitting the 2D DFT matrix of $\left ( 2,2 \right )$ and $\left ( 4,4 \right )$ grid points are $\zeta = 0.8$ and $\zeta = 0.95$, respectively.
 \item A well-trained SIM is capable of generating the angular spectrum of incoming signals and providing a DOA estimate with an MSE of $10^{-4}$ under moderate conditions.
 \item The spacing between adjacent antennas in the receiver array should be no less than $\lambda/2$, while an isomorphic arrangement between the receiver array and input layer achieves the most robust performance.
\end{itemize}

Since this is the first paper on SIM-aided DOA estimation, there remain several open issues that deserve further exploration. Firstly, further investigations are required to leverage SIM technology to realize more advanced DOA estimation algorithms, such as super-resolution and compressed sensing methods in the wave domain. The general scenario of multiple sources should also be considered. Secondly, by integrating amplifiers into each meta-atom and operating them in the non-linear regime, SIMs may be capable of fully realizing DNNs, while using waves for forward computation. This would enable SIMs to process more complex tasks such as near-field positioning. Moreover, the transmission coefficient of each meta-atom may not be continuously controlled in practice. It is crucial to design appropriate algorithms for optimizing the discrete phase shifts and to evaluate the SIM's performance under more realistic response models.

Before concluding, we note that the inverse system of the SIM presented in this paper can be utilized for implementing angular division multiplexing. By employing a SIM as a multi-user precoder at the base station, the signal for each user can be transmitted directly from the corresponding antenna. This would substantially simplify the hardware design of wireless communications. Motivated readers also might like to refer to \cite{JSAC_2023_An_Stacked, ICC_2023_An_Stacked}. In a nutshell, the advanced SIM technology offers a new computational paradigm by directly processing EM waves, which would profoundly influence future system designs for both wireless communication and radar sensing applications.

\begin{appendices}
\section{Proof of \eqref{eq20}}\label{A1}
First, we note that the gradient of the loss function $\mathcal{L}$ \emph{w.r.t.} the phase shift vector of the $l$-th layer $\boldsymbol{\xi} _{l}$ can be expressed as
\begin{align}
\nabla_{\boldsymbol{\xi} _{l}} \mathcal{L} &=\sum_{n=1}^{N}\nabla_{\boldsymbol{\xi} _{l}} \left \|\beta \boldsymbol{g}_{n}-\boldsymbol{f}_{n} \right \|^{2},\ l=1,\cdots ,L. \label{eq43}
\end{align}

Furthermore, the $m$-th entry of the gradient in \eqref{eq43} is obtained by taking the partial derivative of $\left \|\beta \boldsymbol{g}_{n}-\boldsymbol{f}_{n} \right \|^{2}$ \emph{w.r.t.} $\xi _{l,m}$. Applying the chain rule of derivatives yields
\begin{align}\label{eq44}
\frac{\partial \left \|\beta \boldsymbol{g}_{n}-\boldsymbol{f}_{n} \right \|^{2} }{\partial \xi _{l,m}}&=2\Re\left \{ \beta^{\ast } \frac{\partial \boldsymbol{g}_{n}^{H}}{\partial \xi _{l,m}}\left ( \beta \boldsymbol{g}_{n}-\boldsymbol{f}_{n} \right ) \right \} \notag\\
&\overset{\left ( i \right )}{=}2\Re\left \{ \beta^{\ast } \frac{\partial \left (\boldsymbol{P}_{l,n}\boldsymbol{\upsilon} _{l}\right )^{H}}{\partial \xi _{l,m}}\left ( \beta \boldsymbol{g}_{n}-\boldsymbol{f}_{n} \right ) \right \} \notag\\
&=2\Re\left \{ \beta^{\ast } \frac{1}{j}\upsilon _{l,m}^{\ast }\boldsymbol{e}_{m}^{H}\boldsymbol{P}_{l,n}^{H}\left ( \beta \boldsymbol{g}_{n}-\boldsymbol{f}_{n} \right ) \right \} \notag\\
&=2\Im\left \{ \beta^{\ast } \upsilon _{l,m}^{\ast }\boldsymbol{e}_{m}^{H}\boldsymbol{P}_{l,n}^{H}\left ( \beta \boldsymbol{g}_{n}-\boldsymbol{f}_{n} \right ) \right \},
\end{align}
for $m=1,\cdots ,M,\ l=1,\cdots ,L$, where $\left ( i \right )$ holds due to the fact that $\boldsymbol{g}_{n}=\boldsymbol{P}_{l,n}\boldsymbol{\upsilon} _{l}$, and $\boldsymbol{P}_{l,n}$ is defined in \eqref{eq21}, $\boldsymbol{e}_{m}^{H}$ represents the $m$-th row of the identity matrix $\boldsymbol{I}_{M}$.

By gathering the $M$ partial derivatives in \eqref{eq44} into a vector, the gradient in \eqref{eq43} can be obtained by
\begin{align}
 \nabla_{\boldsymbol{\xi} _{l}} \left \|\beta \boldsymbol{g}_{n}-\boldsymbol{f}_{n} \right \|^{2} = 2\Im\left \{ \beta^{\ast } \boldsymbol{\Upsilon} _{l}^{H}\boldsymbol{P}_{l,n}^{H}\left ( \beta \boldsymbol{g}_{n}-\boldsymbol{f}_{n} \right ) \right \}. \label{eq45}
\end{align}

Substituting \eqref{eq45} into \eqref{eq43} completes the proof. $\hfill \square$

\section{Proof of Theorem \ref{theorem1}}\label{A2}
Before proceeding further, we first provide a pair of relevant lemmas.
\begin{lemma}{\emph{(The Three-Moment $\chi ^{2}$ Approximation)}}\label{L1}
Consider a non-singular linear transformation $X$ in the general form of
\begin{align}
 X=\sum_{k=1}^{K}\lambda _{k}\chi _{h_{k}}^{2}\left ( \delta _{k}^{2} \right ), \label{eq46}
\end{align}
where $\lambda _{k}$ are the non-zero coefficients, and $\chi _{h_{k}}^{2}\left ( \delta _{k}^{2} \right ),\ k = 1, \cdots ,K$ represent a set of independent noncentral $\chi ^{2}$ variables with $h_{k}$ DoF and noncentrality parameter $\delta _{k}^{2}$. The three-moment $\chi ^{2}$ approximation for the distribution of $X$ is given by:
\begin{align}
 X\cong \sqrt{\frac{\mu _{2}}{h}} \left ( \chi _{h}^{2}-h \right )+\mu_{1}, \label{eq47}
\end{align}
where we have $h=\mu_{2}^{3}/\mu_{3}^{2}$ such that both sides in \eqref{eq47} have equal third moments \cite{OUP_1959_Pearson_Note}, and $\mu_{i}\triangleq \sum_{k=1}^{K}\lambda _{k}^{i}\left ( h_{k}+i\delta _{k}^{2} \right ),\ i=1,2,3$.

The three-moment approximation can be used when $\mu_{3}>0$ \cite{OUP_1959_Pearson_Note}. Otherwise, the approximation can instead be applied to the distribution of $-X$.
\end{lemma}

\begin{lemma}{\emph{(The Wilson-Hilferty Transformation)}}\label{L2}
For a chi-square variable $X$ with DoF $h$ (i.e., $X\sim \chi_{h}^{2}$), taking the cube root of $X$ divided by $h$ (i.e., $\sqrt[3]{X/h}$) results in a value that is approximately normally distributed with mean of $1-2/\left ( 9h \right )$ and variance of $2/\left ( 9h \right )$ \cite{PNAS_1931_Wilson_The}.
\end{lemma}

Now we continue by deriving an upper bound of the MSE. Note that the $\textrm{MSE} _{\psi_{\text{x}}}$ in \eqref{eq33} and the $\textrm{MSE} _{\psi_{\text{y}}}$ in \eqref{eq34} only depend on the probability of detecting the corresponding index as the peak position, which is defined as
\begin{align}
 \textrm{Pr}\left ( n,t \right )&\triangleq \textrm{Pr}\left \{ \left | r_{n,t} \right |^{2}=\underset{\tilde{n}=1,\cdots ,N,\atop \tilde{t}=1,\cdots , T}{\max} \left | r_{\tilde{n},\tilde{t}} \right |^{2} \right \} \notag\\
 &\leqslant \textrm{Pr}\left \{ \left | r_{n,t} \right |^{2}\geq \left | r_{\breve{n},\breve{t}} \right |^{2} \right \} \notag\\
 &=\textrm{Pr}\left \{ \left | r_{n,t} \right |^{2}-\left | r_{\breve{n},\breve{t}} \right |^{2}\geq 0 \right \}\notag\\
 &=\textrm{Pr}\left \{ D_{n,t}\geq 0 \right \}, \label{eq48}
\end{align}
where we have $D_{n,t} = \left | r_{n,t} \right |^{2}-\left | r_{\breve{n},\breve{t}} \right |^{2}$, and $\left | r_{n,t} \right |^{2}$ and $\left | r_{\breve{n},\breve{t}} \right |^{2}$ are distributed according to the noncentral chi-squared distribution, satisfying
\begin{align}
2\left | r_{n,t} \right |^{2}&\sim {\chi}_{2}^{2}\left ( \left | \sqrt{2\varrho }\tilde{\boldsymbol{g}}_{n}^{H} \boldsymbol{\Upsilon }_{0,t}\boldsymbol{a}\left ( \bar{\psi }_{\textrm{x}},\bar{\psi }_{\textrm{y}} \right )s \right |^{2}\right ),\\
2\left | r_{\breve{n},\breve{t}} \right |^{2}&\sim {\chi}_{2}^{2}\left ( \left | \sqrt{2\varrho }\tilde{\boldsymbol{g}}_{\breve{n}}^{H} \boldsymbol{\Upsilon }_{0,\breve{t}}\boldsymbol{a}\left ( \bar{\psi }_{\textrm{x}},\bar{\psi }_{\textrm{y}} \right )s \right |^{2}\right ).
\end{align}

According to \emph{Lemma \ref{L1}}, the three-moment $\chi ^{2}$ approximation of $2D_{n,t}$ is obtained by
\begin{align}
 2D_{n,t}\cong \sqrt{\frac{\mu _{2,n,t}}{h_{n,t}}} \left ( \chi _{h_{n,t}}^{2}-h_{n,t} \right )+\mu_{1,n,t}, \label{eq49}
\end{align}
where $h_{n,t}$ and $\mu_{i, n,t},\ i=1,2,3$ are defined as in \eqref{eq38} and \eqref{eq40}, respectively.

Hence, the probability $\textrm{Pr}\left \{ D_{n,t}\geq 0 \right \}$ amounts to taking
\begin{align}
\textrm{Pr}\left \{ D_{n,t}\geq 0 \right \}&\cong \textrm{Pr}\left \{ \tilde{D}_{n,t} \geq b_{n,t} \right \}\notag\\
&=\textrm{Pr}\left \{ \sqrt[3]{\frac{\tilde{D}_{n,t}}{h_{n,t}}}\geq \sqrt[3]{\frac{b_{n,t}}{h_{n,t}}} \right \},
\end{align}
where we have $\tilde{D}_{n,t} \sim \chi _{h_{n,t}}^{2}$ and $b_{n,t}$ is defined in \eqref{eq39}.

Upon applying the Wilson-Hilferty approximation in \emph{Lemma \ref{L2}}, we have
\begin{align}
 \sqrt[3]{\frac{\tilde{D}_{n,t}}{h_{n,t}}} \sim \mathcal{N}\left ( 1-\frac{2}{9h_{n,t}},\frac{2}{9h_{n,t}} \right ).
\end{align}

Therefore, we arrive at
\begin{align}
 &\textrm{Pr}\left \{ \sqrt[3]{\tilde{D}_{n,t}/h_{n,t}}\geq \sqrt[3]{b_{n,t}/h_{n,t}} \right \} \notag \\
 =&Q\left ( \frac{9h_{n,t}}{2}\sqrt[3]{\frac{b_{n,t}}{h_{n,t}}}-\frac{9h_{n,t}}{2}+1 \right ). \label{eq52}
\end{align}

Upon substituting \eqref{eq46} and \eqref{eq52} into \eqref{eq33} and \eqref{eq34}, the proof is completed. $\hfill \square$
\end{appendices}
\bibliography{ref}
\bibliographystyle{IEEEtran}
\end{document}